\documentclass[twocolumn]{aastex62}

\usepackage{amsmath, amsthm, amssymb,eufrak}
\usepackage{graphicx}

\usepackage{soul}
\usepackage{color}
\usepackage{dcolumn}% Align table columns on decimal point
\usepackage{bm}% bold math
\usepackage[mathscr]{euscript}
\usepackage{mathrsfs}
\usepackage{amsbsy}
\begin{document}

\title{\bf Modified Friedmann equations via conformal Bohm -- De Broglie gravity}

\correspondingauthor{G. Gregori}
\email{gianluca.gregori@physics.ox.ac.uk}

\author{G. Gregori}
\affiliation{Department of Physics, University of Oxford, Parks Road, Oxford OX1 3PU, UK}

\author{B. Reville}
\affiliation{Max-Planck-Institut f\"ur Kernphysik, Postfach 10 39 80, 69029 Heidelberg, Germany}
\author{B. Larder}
\affiliation{Department of Physics, University of Oxford, Parks Road, Oxford OX1 3PU, UK}

%\author{S. Sarkar}
%\affiliation{Department of Physics, University of Oxford, Parks Road, Oxford OX1 3PU, UK}
%\affiliation{Niels Bohr Institute, Blegdamsvej 17, 2100 Copenhagen, Denmark}

\begin{abstract}
We use an alternative interpretation of quantum mechanics, based on the Bohmian trajectory approach, and show that the quantum effects can be included in the classical equation of motion via a conformal transformation on the background metric. We apply this method to the Robertson-Walker metric to derive a modified version of Friedmann's equations for a Universe consisting of scalar, spin-zero, massive particles. These modified equations include additional terms that result from the non-local nature of matter and appear as an acceleration in the expansion of the Universe. We see that the same effect may also be present in the case of an inhomogeneous expansion.	 
\end{abstract}


\section{Introduction}
\noindent
The equations governing quantum mechanics have been known for nearly
100 years, but yet the task of reconciling them with the equations of classical motions remains unsolved (see e.g., Ref. \cite{hiley1995the}). The problem is usually expressed in terms of the trajectory (or path) that a particle follows. While in ordinary Newtonian theory a particle moves along a well defined path (a concept that can be extended to curved space too), this is not true anymore in orthodox (or the Copenhagen interpretation) quantum mechanics. This is because it is 
impossible to simultaneously measure the position and momentum of the particle.
There is, however, an alternative possibility, called Bohmian mechanics or the de Broglie-Bohm interpretation \citep{bohm1952a,bohm1987an}, whereby the particle follows a definite path, different from the classical one, determined not only by the action of local potential, as in ordinary classical mechanics, but also by the non-local feedback of the particle's wave function on its own motion \citep{bohm1952a}.   

It has been shown, in the non-relativistic case, that Bohmian quantum mechanics yields the same results as orthodox quantum mechanics \citep{bohm1952a}, indeed both approaches can be developed from the usual Schr\"{o}dinger equation, albeit with a different physical interpretation. Experiments have also demonstrated that the de Broglie-Bohm trajectories can have a sound physical interpretation if full non-locality is accounted for \citep{mahler2016experimental}. Extensions to quantum fluids \citep{HaasQuantumPlasmas,cross}, field theory \citep{drr2004bohmian,carroll2007metric} and curved space-time \citep{drr2014can,holland1992the,shojai2006about} have also been presented.

Here we adopt one of the latter approaches (see \cite{carroll2005fluctuations} for a review) and develop a formalism which permits a reinterpretation of the classical path by including the effects of Bohm's nonlocal potential on the particle's motion. 
\color{black}
%Since most of these techniques are unfamiliar in orthodox quantum mechanics, 
This is achieved in the usual manner, through suitable modification of the action for a free particle of mass $m$,
\begin{equation}
\label{EMaction}
\mathcal{S}=-m\int  \sqrt{g_{\mu \nu} u^\mu u^\nu} ds 
\end{equation}
where $u^\mu=d x^\mu/ds$ is the dimensionless 4-momentum and $g_{\mu\nu}$ the space-time metric (we use units, unless specified otherwise, where $c=\hbar=1$). 
\color{black}
The metric is taken to have signature $(+,-,-,-)$. 

The effects of an external field can be incorporated into the above action via a modified metric: ${g}_{\mu \nu}\rightarrow \tilde{g}_{\mu \nu}$. As discussed in \cite{goldstein1950}, there is no single way to modify the metric in equation (\ref{EMaction}), and in fact, any function of a world scalar (in this case $u_\mu u^\mu$) that leads to the correct equations of motion is acceptable. 
The same is also true for actions that account for the particle's interaction with external forces by including an additive term, provided again due care is taken to ensure Lorentz invariance. 
\cite{zee2013}, for example, exploits a number of different approaches to introduce the effects of electromagnetic and gravitational fields into the action. As an example, it is well known that a simple modification of the Minkowski metric tensor: \begin{equation}
{\eta}_{\mu \nu}\rightarrow \tilde{\eta}_{\mu \nu}
= {\rm diag}(1+2V/m,-1,-1,-1),
\label{g1}
\end{equation}
for a scalar potential $V$,
provides a way to reproduce Newton's law in the non-relativistic limit. 
We will employ a similar approach below in order to introduce quantum effects into more complex metrics.

\color{black}
\section{Quantum effects as a conformal transformation of the metric}
\noindent
%\cite{bohm1952a}.
%Starting with the simple case of a non-relativistic spinless particle of mass $m$ in an external potential $V_{\rm ext}$, in the Bohm picture, the particle trajectory is governed by the wavefunction satisfying Schr\"odinger's equation 

Writing  Schr\"odinger's equation, for a particle of mass $m$, in polar form: ${\rm \psi}= \left|\psi\right| \exp\left({\rm i}S\right)$ with real function $S(\bm{x},t)$, \cite{bohm1952a} establishes the set of coupled equations 
\begin{equation}
\partial_t |\psi|^2 + \bm{\nabla}\cdot\left(\vec{v}\, |\psi|^2\right) = 0
\end{equation}
and
\begin{equation}
\label{HJ}
\partial_t S + \frac{1}{2m} (\nabla S)^2-\frac{1}{2 m} \frac{\nabla^2 |\psi|}{|\psi|}+V_{\rm ext} = 0,
\end{equation}
where we have set $\vec{p}  \equiv \bm{\nabla} S = m \vec{v} = m\,d \vec{x}/dt$, and $V_{\rm ext}$ is the external potential acting on the particle. 
The former equation corresponds to conservation of probability density, while the latter is recognised as the Hamilton-Jacobi equation. The analogy with classical mechanics is self-evident on defining the so-called Bohm potential
\begin{equation}
V_B=-\frac{1}{2 m} \frac{\nabla^2 |\psi|}{|\psi|}.
\end{equation}
%Bohm's original derivation \cite{bohm1952a} (see also Ref.~\cite{durr} for further discussions), Schr\"odinger's equation
%\begin{equation}
%{\rm i} \partial_t \psi = -\frac{1}{2m}\nabla^2
%\psi + V_{\rm ext} \psi \enspace ,
%\end{equation}
%is written in polar form ${\rm \psi}= \left|\psi\right| \exp\left({\rm i}S\right)$, with real function $S(\bm{x},t)$. 
%Equating real and imaginary parts results in a continuity equation equivalent to the conservation of probability density $\partial_t |\psi|^2 + \bm{\nabla}\cdot\left(\vec{v}\, |\psi|^2\right) = 0$, while $S$ can be shown to satisfy the Hamilton-Jacobi equation
%\begin{equation}
%\label{HJ}
%\partial_t S + \frac{1}{2m} (\nabla S)^2+V_{\rm ext}+V_B = 0,
%\end{equation}
%where as with its classical counterpart
%$\vec{p} = m \vec{v} = \bm{\nabla} S$.
%The analogy with classical mechanics is self-evident having defined the so-called Bohm potential
%\begin{equation}
%V_B=-\frac{1}{2 m} \frac{\nabla^2 |\psi|}{|\psi|}.
%\end{equation}
Thus, in Bohm's interpretation, the particle moves as guided by its own wave function. In the limit of $V_B \ll V_{\rm ext}$ the trajectory is indistinguishable from its classical one.

As mentioned earlier, an alternative viewpoint is to rewrite (\ref{EMaction}) in the Minkowski metric as
\begin{eqnarray}
\mathcal{S}&=&-m\int\sqrt{\tilde{g}_{\mu \nu} u^\mu u^\nu} ds,
\end{eqnarray}
where following \cite{zee2013} we set
\begin{equation}
\tilde{g}_{\mu \nu} = \left(1+\frac{2V}{m}\right) \eta_{\mu \nu}.
\label{g3}
\end{equation}

%As an illustrative example, we consider the non-relativistic ($|d\vec{x}/dt| \ll 1$, and $m \gg V$) limit, where 
%\begin{equation}
%\tilde{g}_{\mu \nu} \approx \left(1+\frac{2V}{m}\right) \eta_{\mu \nu},
%\label{g}
%\end{equation}
%$V \equiv V(x^\mu)$ is the total potential. For a vanishing variation, it follows that
%\begin{eqnarray}
%\delta \mathcal{A} = -m \delta \int dt 
%\sqrt{
%\left(1+\frac{2V}{m}\right)
%\bigg(1-\left|\frac{d\vec{x}}{dt}\right|^2\bigg)
%}=0.
%\end{eqnarray}

To order $|d\vec{x}/dt|^2$, ignoring additive constants, the above action gives the classical Lagrangian
\begin{equation}
\label{LClassic}
L \simeq \frac{m}{2} \left(\frac{d\vec{x}}{dt}\right)^2-V,
\end{equation}

with associated Hamiltonian
\begin{eqnarray}
\mathcal{H} = \frac{m}{2}\left(\frac{d\vec{x}}{dt}\right)^2+V = \frac{|\vec{p}|^2}{2 m}+V.
\label{energy}
\end{eqnarray}

%So far, this is nothing new -- we have derived the equation of motion for a purely classical particle.

\hspace{0.1in}

Exploiting our prior knowledge of the Bohm potential, i.e., letting $ V = V_{\rm ext} + V_{\rm B}$, and carrying out the standard substitution $ \mathcal{H}= -\partial_t S $
it is immediately observed that (\ref{energy}) reproduces (\ref{HJ}) as expected. \color{black}
Thus, if we acknowledge the presence of the Bohm potential in a given system, it is possible to include its effect in macroscopic systems via an appropriate transformation of the metric. Recent work that exploits this approach \citep{carroll2005fluctuations}, tacitly assumes the equations already describes an ensemble or cloud of particles. We attempt to make this concept more precise in the next section.

\color{black}
%With this substitution, the energy equation (\ref{energy}) reduces to the well-known Hamilton-Jacobi equation of classical mechanics
%\begin{equation}
%\partial_t S + \frac{1}{2m} (\partial_a S)^2+V = 0 \enspace .
%\end{equation}
%Combining this equation with the necessary requirement of conservation of probability density 
%\begin{eqnarray}
%\partial_t |\psi|^2 + \partial_a\left(v^a |\psi|^2\right) = 0,
%\end{eqnarray}
%where $v^a = p^a/m$, Schr\"{o}dinger's equation can be reconstructed. 
%The case of a particle motion subject to an external (local) potential, $V_{\rm ext}$, is straightforward. This is done by employing a conformal transformation on the Minkowski metric of the form of Eq. \ref{g}, but with $V=V_B+V_{\rm ext}$.

\section{The many-body version of the Bohm potential}
\noindent
\color{black}
While, so far, we have considered the motion of a single test particle, it is clear that we cannot neglect the effects of all the other particles in the system. A full theoretical description of this quantum system must take into account effects such as non-locality and correlations on all the scales of interest. A continuum description of such quantum gas is often more convenient, and it usually done in terms of an
average density, by replacing the product of many-body wavefunctions with an appropriate density operator into the relevant equations of motion. 
In previous work \citep{carroll2005fluctuations}, the transition from a microscopic to a continuum (macroscopic) description of the Bohm's potential was implicitly assumed.

In order to justify a continuum approach in the Bohm picture more rigorously, let us consider an isolated quantum system of $N$ particles of mass $m$, described by
a many-body wavefunction 
$\psi=\psi(x_1,x_2,...,x_N)$, where $x_i$ is the spatial coordinate of the $i^{\rm th}$ particle.
For simplicity, in the following, we take $V_{\rm ext}=0$.
%We will show that the resultant contribution of the environment can also be written in terms of a de Broglie-Bohm potential, where the square of the single particle wavefunction is replaced by the density distribution.   
\color{black} The $N$-body Bohm potential can be written as \citep{bohm1952a}:
\begin{eqnarray}
V^{(N)} = -\frac{1}{2 m} \sum_{i=1,N} \frac{\nabla_i^2 |\psi|}{|\psi|},
\end{eqnarray}
where $\nabla_i$ is the gradient with respect to the $i^{\rm th}$ particle coordinates. It is first assumed that the many-body wavefunction can be written as a product of single-particle wavefunctions: $\psi(x_1,x_2,...,x_N)=\varphi_1(x_1)\varphi_2(x_2)...\varphi_N(x_N)$, where $\varphi_i(x_i)$ is the $i^{\rm th}$ particle wavefunction. The expectation value of the Bohm potential is given by  
\begin{widetext}
\begin{eqnarray}
\langle V \rangle &=& \int dx_1 \int dx_2 ... \int dx_N V^{(N)} |\psi|^2 
= \sum_{i=1,N} \int dx_i V^{(N)}_i(x_i) |\varphi_i(x_i)|^2 
= \int dx \sum_{i=1,N} V^{(N)}_i(x) |\varphi_i(x)|^2,
\label{mb1}
\end{eqnarray}
\end{widetext}
where,
\begin{equation}
    V^{(N)}_i(x) = -\frac{1}{2 m} \frac{\nabla^2 |\varphi_i(x)|}{|\varphi_i(x)|}
\end{equation}
is the single-particle Bohm potential. The total particle density is $n(x) = \sum_i |\varphi_i|^2 = \sum_i n_i$, where $n_i$ is the probability distribution for the $i^{\rm th}$ particle.
This gives
\begin{widetext}
\begin{eqnarray}
\langle V \rangle = \int dx \sum_{i=1,N} n_i \left( -\frac{1}{2 m} \frac{\nabla^2 \sqrt{n_i}}{\sqrt{n_i}} \right) \approx \int dx\, n(x) \left( -\frac{1}{2 m} \frac{\nabla^2 \sqrt{n}}{\sqrt{n}} \right),
\label{mb2}
\end{eqnarray}
\end{widetext}
where the last step applies when each particle's wavefunction significantly extends over their inter-particle separation, and so the single-particle's amplitudes are assumed to be the same throughout space (that is, $|\varphi_i(x)| \equiv  |\varphi_0(x)|$).
This is the linearization approximation of quantum hydrodynamics \citep{manfredi2001selfconsistent,graziani,Moldabekov},
\color{black}
and it strictly applies when the number density of particles varies smoothly in space. 
However, it can also be shown \citep{graziani,Moldabekov} that even in presence of strong inter-particle correlations, equation (\ref{mb2}) continues to apply approximately, to order ${\cal O}(1)$. 
\color{black}
By comparing equations (\ref{mb1}) and (\ref{mb2}) we see that the many-body Bohm potential can be thus written as a function of a single spatial coordinate as
\begin{equation}
    V(x) = -\frac{1}{2 m} \frac{\nabla^2 n^{1/2}}{n^{1/2}}.
\end{equation}
\color{black}
The many-body quantum potential discussed here and its application to extended systems of electrons and ions has recently 
been tested against more established solutions of the Sch\"odinger's equation based on density function theory (DFT) \citep{Hohenberg1964,Kohn1965}. In fact, the quantum potential approach shows excellent agreement with DFT, with substantial reduction in computational speed \citep{2018arXiv181108161L}.
\color{black}

\color{black}
The complexity of quantum mechanics is not circumvented in Bohm's interpretation; the particles' trajectories are determined by this nonlocal potential, which depends on the particles' wave function itself. However, what Bohm's description offers is a recipe that allows us to obtain quantum corrections to the classical equation.  

\section{The relativistic version of the Bohm potential}
\noindent
\color{black}
Extensions to cosmological models also requires a relativistic and covariant treatment for relevant dark matter candidates. With a view to exploring the Bohm potential associated with such dark matter candidates, we apply the previous analysis to the Klein-Gordon equation \citep{nikoli2005relativistic,nikoli2007bohmian,shojai2006about,carroll2005fluctuations}
\begin{equation}
\Box~\psi +m^2 \psi = 0 \, 
\end{equation}
where $\Box = g^{\mu\nu}\nabla_\mu\nabla_\nu$ is the d'Alembertian, and
$\nabla_\mu$ the covariant derivative with respect to the metric $g_{\mu \nu}$. In the Minkowski metric $\Box = \partial_t^2 - \nabla^2$.

Following \cite{deBroglie}, writing the Klein-Gordon equation in polar form, and separating out the real part gives
\begin{eqnarray}
g^{\mu\nu} \nabla_\mu S \,\nabla_\nu S = \left(1+\frac{2V_B}{m}\right)m^2 \equiv \mathcal{M}^2 \, ,
\label{mm}
\end{eqnarray}
where $V_B$ is now in covariant form
\begin{eqnarray}
V_B = \frac{1}{2 m} \frac{\Box\, |\psi|}{|\psi|}\,.
\label{vg}
\end{eqnarray}
The question, however, arises as Eq.~(\ref{mm}) can not be adopted as the relativistic
Hamilton-Jacobi equation \citep{holland1992the,carroll2005fluctuations}. The fact that $\mathcal{M}$ is not positive definite is a serious concern, and furthermore ($\ref{mm}$) leads to tachyonic solutions, which we must exclude on physical grounds -- although tachyonic Bohm theories have been discussed in the literature \citep{gonzlezdaz2004subquantum}.
The correct approach is that suggested by
\cite{shojai2006about}, giving
\begin{equation}
\mathcal{M}^2 = m^2 ~{\rm exp}\left(\frac{2V_B}{m}\right)~.
\label{qmexp}
\end{equation}
This reduces to the previous case in the weak field limit. 

\color{black}
To see better how the exponential form for the effective mass, $\mathcal{M}$, arises, we follow the approach of \cite{shojai2006about}.
By minimizing the relativistic action, we easily obtain the equation of motion for a particle of variable mass $\mathcal{M}=\mathcal{M}(x)$ in the Minkowski metric,
\begin{equation}
    \mathcal{M} \frac{d u_\mu}{d \tau} = \left(\eta_{\mu \nu} - u_\mu u_\nu \right)\partial^\nu \mathcal{M}, 
\end{equation}
where $d\tau^2 = dx_\mu dx^\mu$ is the proper time. By taking the non-relativistic limit (i.e., assuming small velocities and $t=\tau$), the space components of the equation of motion can be written as
\begin{equation}
  \mathcal{M} \frac{d^2 \vec{x}}{d t^2} = -\nabla \mathcal{M},
\end{equation}
or, alternatively, as
\begin{equation}
  m \frac{d^2 \vec{x}}{d t^2} = -\nabla \left(m \ln \frac{\mathcal{M}}{\mathfrak{m}}\right),
  \label{qmeff}
\end{equation}
with $m$ the bare mass, and $\mathfrak{m}$ is some arbitrary mass scale. Now, if we require that eq.~(\ref{qmeff}) describes the motion of a quantum particle of mass $m$, then consistency with the Hamiltonian (\ref{energy}), implies
\begin{equation}
   \mathcal{M} = \mathfrak{m} \exp\left( -\frac{1}{2 m^2} \frac{\nabla^2 |\psi|}{|\psi|} \right).
\end{equation}
The relativistic generalization would then lead to
\begin{equation}
   \mathcal{M} = \mathfrak{m} \exp\left(\frac{1}{2 m^2} \frac{\Box^2 |\psi|}{|\psi|} \right),
\end{equation}
which is the same as (\ref{qmexp}) if we choose the mass scale $\mathfrak{m} \equiv m$.

\color{black} As further pointed out by \cite{deBroglie} (chap. 10), the inclusion of quantum effects are entirely 
equivalent to the change of the space-time metric
$g_{\mu\nu}\rightarrow \tilde{g}_{\mu\nu} = (\mathcal{M}^2/m^2) g_{\mu\nu}$,
that is, a conformal transformation. The approach of \cite{shojai2006about} generalises this result to ensure it satisfies the correct non-relativistic limit.
%Rewriting the Hamilton-Jacobi equation as
%\begin{eqnarray}
%(m^2/\mathcal{M}^2)\, g^{\mu\nu} \nabla_\mu S \,\nabla_\nu S = m^2,
%\label{mm2}
%\end{eqnarray}
%we immediately see that this is equivalent to a change of the metric
%$g^{\mu\nu}\rightarrow \tilde{g}^{\mu\nu} = (\mathcal{M}^2/m^2) g^{\mu\nu}$,
%that is, a conformal transformation. 
Identifying the four momentum $P_\mu =\nabla_\mu S $,
the relativistic energy equation $g^{\mu\nu} P_\mu P_\nu = m^2$ is thus recovered as the weak quantum potential limit of a more general equation $\tilde{g}^{\mu\nu} \tilde{\nabla}_\mu S \, \tilde{\nabla}_\nu S=
m^2$, where $\tilde{\nabla}_\mu$ is the covariant derivative with respect to the metric $\tilde{g}_{\mu\nu}$.  
\color{black}

Semi-relativistic approaches, where the quantum potential is simply added to the equation of motions without a conformal transformation, have also been studied \citep{das2014quantum}, and lead to solutions that are different from standard cosmology \citep{ali2015cosmology}.

%A possible solution to the tachionic problem consists in making the \emph{ansatz}

Following the same arguments as earlier, we can extend this previous result to the many-body system, such that the conformal transformation we should apply has the form
\begin{eqnarray}
 \tilde{g}_{\mu\nu} = {\rm e}^Q \, g_{\mu\nu},
\end{eqnarray}
 where
\begin{eqnarray}
Q = \frac{1}{m^2} \frac{\Box\, n^{1/2}}{n^{1/2}}.
\label{grad}
\end{eqnarray}

\color{black}
Thus, given $Q$, quantum effects on classical bodies moving in space-time can be calculated via this prescription. In the next section, this is applied to the Einstein field equations.

\section{Extension to cosmology}
\noindent
%We now postulate that the Bohm trajectory approach applies not only to a Minkowski space-time, but to any general space-time with background metric $g_{\mu \nu}$ \cite{nikoli2005relativistic,nikoli2007bohmian,shojai2006about,carroll2005fluctuations}.
%Thus, quantum effects on classical bodies moving in this space-time can be calculated via a conformal transformation
%\begin{eqnarray}
%\tilde{g}_{\mu \nu} = \left(1+\frac{2 V}{m}\right) g_{\mu \nu},
%\label{gg}
%\end{eqnarray}
%with
%\begin{eqnarray}
%V = \frac{1}{2 m} \frac{\Box\, n^{1/2}}{n^{1/2}},
%\label{vg}
%\end{eqnarray}
%where $\Box = g^{\mu\nu}\nabla_\mu\nabla_\nu$ is the d'Alembertian, with
%$\nabla_\mu$ being the covariant derivative with respect to the metric $g_{\mu \nu}$. In the Minkowski
%metric we would have $\Box = \partial_t^2 - \nabla^2$. 
%\color{blue}
%The above form follows directly from consideration of the polar treatment of the wavefunction applied to the Klein-Gordon equation
%\begin{equation}
%\Box~\psi +m^2 \psi = 0 \, .
%\end{equation}
%resulting in 
%\begin{eqnarray}
%g^{\mu\nu} \nabla_\mu S \,\nabla_\nu S = \left(1+\frac{2V}{m}\right)m^2 \equiv \mathcal{M}^2
%\end{eqnarray}
%It follows that the above conformal transformation (\ref{gg}) is equivalent to a mass renormalisation in the relativistic energy equation. Thus, provided one remains consistent in their definition of momenta and energy operators relative to the mass $\mathcal{M}$, the standard results follow.
\color{black}
%There are however, some issues in the above approach \cite{shojai2006about,holland1992the}. This is because the Bohm potential can assume both positive and negative values, and so, for large negative $V/m$, the overall sign of the metric can change. This introduces tachionic solutions \cite{shojai2006about}, which we must exclude on physical grounds -- althought tachionic Bohm theories have been discussed in the literature \cite{gonzlezdaz2004subquantum}. Alternatively, semi-relativistic approaches, where the quantum potential is simply added to the equation of motions without a conformal transformation, have also been studied \cite{das2014quantum}, and lead to solutions that are different from standard cosmology \cite{ali2015cosmology}.

%A possible solution to the tachionic problem consists in making the \emph{ansatz} that Eq. (\ref{vg}) is the weak potential limit of \tilde{g}_{\mu \nu} = e^Q %g_{\mu \nu} = \lambda^2 g_{\mu \nu}$ \cite{shojai2006about,carroll2005fluctuations}, where
%\begin{eqnarray}
%Q = \frac{1}{m^2} \frac{\Box\, n^{1/2}}{n^{1/2}}.
%\label{grad}
%\end{eqnarray}
%Clearly, in the non-relativistic limit $m \gg V$ and so $Q\ll 1$, which reproduces the earlier results to the same order.

We proceed by setting $g_{\mu \nu}$ to be the Robertson-Walker metric of an expanding Universe. The line element in such a metric is given by
\begin{widetext}
\begin{eqnarray}
ds^2 = dt^2-R^2(t) \left[ \frac{dr^2}{1-\kappa r^2}+r^2 (d\theta^2+\sin^2\theta \, d\phi^2) \right] \equiv dt^2 - R^2(t)\, d\Omega^2,
\end{eqnarray}
\end{widetext}
where $R(t)$ is the scale factor, set to unity at present time, and $\kappa$ is the curvature parameter. This allows some simple scaling relations for the quantum potential $Q$ to be obtained. In the homogeneous approximation, the density is a function of time only. For the Robertson-Walker metric the d'Alembertian operator is 
\begin{eqnarray}
\Box\, n^{1/2} = \frac{1}{R^3} \frac{\partial}{\partial t} \left( R^3 \frac{\partial}{\partial t} \right) n^{1/2}\enspace .
\end{eqnarray}
Assuming a matter-like field $n \sim 1/R^3$, the quantum potential has the form 
\begin{eqnarray}
Q=-\frac{3}{2}\left(\ddot{R}R/\dot{R}^2+\frac{1}{2}\right)H^2/m^2\enspace,
\label{Qdefinition}
\end{eqnarray}
where one can identify $q_{\rm d}=-\ddot{R}R/\dot{R}^2$ as the usual deceleration parameter and $H=\dot{R}/R$ is the Hubble parameter.
%Taking the Hubble parameter $H=\dot{R}/R = \alpha/t$ where $\alpha$ is a constant, we obtain $Q=\left(\frac{3}{2\alpha}-\frac{9}{4} \right) H^2/m^2 \equiv \gamma H^2/m^2$.
%In ordinary cosmology, the solution of Einstein's field equations in the Robertson-Walker metric leads to Friedmann's equations, which predict $H=\dot{R}/R = \alpha/t$, where $\alpha=1/2$ for a radiation dominated universe and $\alpha=2/3$ for a cold matter dominated universe. This implies that $\gamma=3/4$ for the former, and $\gamma=0$ for the latter. In the general case, we expect $\gamma >0$ for the whole evolution of the Universe. Similarly, for a weak quantum potential, $\lambda \sim 1+\alpha^2 \gamma/2 m^2 t^2$. 

%In ordinary cosmology, the solution of Einstein's field equations in the Robertson-Walker metric leads to Friedmann's equations, which predict for both radiation dominated and non-relativistic matter the relation $H \sim \alpha/t$ (where $\alpha=1/2$ for radiation dominated and $\alpha=2/3$ for non-relativistic matter); thus 
%$Q \sim -2\alpha^2/m^2 t^2$, and 

We now consider the conformal
transformation $\tilde{g}_{\mu \nu}={\rm e}^Q ~g_{\mu\nu}$ on the
Robertson-Walker metric.
From the above consideration, we see that the conformal factor has a non-trivial dependence on time.
In fact, its
scaling with time has been discussed in the literature \citep{canuto1977scalecovariant,maeder2017an} -- while it is required by all scale-invariant gravity theories, it has been only justified via arguments needed to match the present value for the cosmological constant \citep{canuto1977the}.
The modified Einstein field equation now reads as \citep{canuto1977scalecovariant,maeder2017an}
\begin{equation}
    \tilde{\cal R}_{\mu\nu} - \frac{1}{2} \tilde{g}_{\mu \nu} \tilde{\cal R} = 8\pi G \tilde{\cal T}_{\mu \nu} + \Lambda \tilde{g}_{\mu \nu},
    \label{eeq}
\end{equation}
where $\tilde{\cal R}_{\mu\nu}$ is the Ricci tensor with respect to the modified metric $\tilde{g}_{\mu \nu}$, $\tilde{\cal R}$ is the Ricci scalar, $\tilde{\cal T}_{\mu \nu}$ is the energy-momentum tensor, and $\Lambda$ is the effective cosmological constant \citep{weinberg1989}.
\color{black}
Since the left hand side of eq. (\ref{eeq}) is manifestly scale-invariant, the same must be true for the right hand side,  in particular 
the energy-momentum tensor   \citep{canuto1977scalecovariant,maeder2017an}, that is, $\tilde{\cal T}_{\mu \nu}={\cal T}_{\mu \nu}$. \color{black} The stress tensor, in general, contains two terms \citep{carroll2007metric}
$\tilde{\cal T}_{\mu \nu} = \tilde{\cal T}_{\mu \nu}^{(M)}+\tilde{\cal T}_{\mu \nu}^{(Q)}$. \color{black} The first one is related to the matter contribution, ${\cal T}_{\mu \nu}^{(M)} = (p+\rho)u_\mu u_\nu -p g_{\mu \nu} $ (for a perfect fluid), where $p$ is the pressure and $\rho$ is the energy density of the matter. A consequence of the scale invariance of the energy-momentum tensor is that both the pressure and energy density of matter are not scale invariant, as discussed in \cite{maeder2017an}.

\color{black} The second term of the energy-momentum tensor, $\tilde{\cal T}_{\mu \nu}^{(Q)}$, instead, arises from the energy density of the quantum potential. This is a quantum gravity modification, implying that quantum effects back react on the metric. As these terms appear as corrections of order ${\cal O}(m^2/m_P^2)$ \citep{pintoneto2018}, where $m_P = 2.4 \times 10^{18}$ GeV is the reduced Planck mass, we can ignore them for particle masses much smaller than $m_P$.
\color{black}

The effective cosmological constant contains two terms, $\Lambda=\Lambda_E+
8\pi G \langle \rho_{\rm vac} \rangle$.
The first one is a constant. The second term, instead, is the vacuum energy contribution that arise from field theory. \color{black}
%In the Bohmian picture the energy density of the vacuum in a mini-superspace with a canonical scalar field can be written as \cite{pintoneto2018}
%\begin{equation}
% \rho_{\rm %vac}=\frac{\dot\phi^2}{2{\cal %N}^2}+V_M,   
%\end{equation}
%where $\phi$ is the scalar field, $V_M$ its potential and $\cal N$ the lapse function.
Estimating the actual value for $\langle \rho_{\rm vac} \rangle$ requires further approximations. In Minkowski space time, such estimate can be obtained by
summing up the zero point contributions for each normal mode of a scalar field of mass $m$; we have \citep{weinberg1989},
\begin{equation}
\langle \rho_{\rm vac} \rangle = \int_0^{k_c} \frac{4 \pi k^2 dk}{(2 \pi)^3} \frac{1}{2}
\sqrt{k^2+m^2} \approx \frac{k_c^4}{16 \pi^2},
\end{equation}
with $k_c \gg m$ some wavelength cut-off which is usually taken to be set by the reduced Planck length, that is $k_c=1/\ell_P = (8 \pi G)^{-1/2}$. Since cosmological observations tells us that $\Lambda \ll 8\pi G \langle \rho_{\rm vac} \rangle$ \citep{weinberg1989}, then it must be that $\Lambda_E$ and $8\pi G \langle \rho_{\rm vac} \rangle$ cancels out each other to a very high degree of accuracy, and, in fact, $\Lambda = 0$ may be a good approximation \citep{weinberg1989,sahni2008republication,coleman1988why,hawking1984the}. The latter statement can be justified from a thermodynamic argument \citep{volovik2006from,bucher1999is}. If we consider a vacuum Universe, the total pressure is $p=-\Lambda$. Interestingly, the same relation applies for a quantum fluid at zero temperature, i.e., in the absence of any finite temperature excitations. If there are no external forces acting on the fluid, then it must be $p=0$ and so $\Lambda=0$ \citep{volovik2001superfluid,volovik2006from}. While $\Lambda=0$ is still hypothetical, it remains plausible in different approaches of quantum gravity \citep{coleman1988why,hawking1984the,barrow2011new,ng1990possible}.

The solution of Einstein field equations (\ref{eeq}) gives modified Friedmann equations \citep{canuto1977scalecovariant,maeder2017an}:
\begin{widetext}
\begin{eqnarray}
\frac{8 \pi G}{3} \rho &=& \frac{\kappa}{R^2}+\frac{\dot{R}^2}{R^2} +
2 \frac{\dot{\lambda} \dot{R}}{\lambda R} + \frac{\dot{\lambda}^2}{\lambda^2}- \frac{{\Lambda}\lambda^2}{3}\enspace,
\label{f1} \\
-{8 \pi G}p&=& \frac{\kappa}{R^2}+2\frac{\ddot{R}}{R}+\frac{\dot{R}^2}{R^2} + 2\frac{\ddot{\lambda}}{\lambda}  
+4 \frac{\dot{\lambda} \dot{R}}{\lambda R} - \frac{\dot{\lambda}^2}{\lambda^2}
-{\Lambda}\lambda^2 ,
\label{f2}
\end{eqnarray}
\end{widetext}
where $\lambda^2 = {\rm e}^Q$.
It is easy to see that the modified Friedmann's equations, (\ref{f1}) and (\ref{f2}), reduce to the usual ones if $\lambda=1$. 
This is perhaps more transparent if we combine the equations to read
%If the conformal factor varies as $\lambda \sim 1+\alpha^2/m^2 t^2$, as discussed earlier, then $2 \ddot{\lambda}/\lambda \simeq 3 \dot{\lambda}^2/\lambda^2$, which simplifies Friedmann's equation to
%\begin{eqnarray}
%\frac{\dot{R}^2}{R^2}=\frac{8 \pi G}{3} \rho + \frac{\tilde{\Lambda}}{3} - \left(
%2 \frac{\dot{\lambda} \dot{R}}{\lambda R} + 
%\frac{\dot{\lambda}^2}{\lambda^2} \right)
%-\frac{k}{R^2},
%\label{f1}
%\end{eqnarray}
%\begin{eqnarray}
%\frac{\ddot{R}}{R}+\frac{\dot{R}^2}{2 R^2}=-4 \pi G p + \frac{\tilde{\Lambda}}{2} - \left(
%2 \frac{\dot{\lambda} \dot{R}}{\lambda R} + 
%\frac{\dot{\lambda}^2}{\lambda^2} \right) -\frac{k}{2 R^2},
%\label{f2}
%\end{eqnarray}
\begin{eqnarray}
\frac{\dot{R}^2}{R^2}=\frac{8 \pi G}{3} \rho + \frac{\tilde{\Lambda}}{3} -
2 \frac{\dot{\lambda} \dot{R}}{\lambda R} - \frac{\dot{\lambda}^2}{\lambda^2}
-\frac{\kappa}{R^2},
\label{f1a}
\end{eqnarray}
\begin{eqnarray}
\frac{\ddot{R}}{R}=-\frac{4 \pi G}{3}(\rho+3p) + \frac{\tilde{\Lambda}}{3} - 
\frac{\dot{\lambda} \dot{R}}{\lambda R} + \frac{\dot{\lambda}^2}{\lambda^2}
-\frac{\ddot{\lambda}}{\lambda},
\label{f2a}
\end{eqnarray}
with $\tilde{\Lambda}=\Lambda \lambda^2$.

\color{black}We notice that similar equations also appear in quantum gravity theories where the infinite degrees of freedom associated to the solution of the Wheeler-DeWitt equation are collapsed under symmetry constraints, a so called gravity-matter mini-superspace \citep{vink1992,pintoneto2018}. The simplest mini-superspace model is described by the metric \color{black}
\begin{equation}
    ds^2 = {\cal N}^2(t)\,dt^2 - R^2(t)\, d\Omega^2,
    \label{msp}
\end{equation}
where ${\cal N}(t)$ is the lapse function (which sets the time gauge).
Quantum effects are then introduced via the Wheeler-DeWitt equation, leading to a modified equation of motion of the form \citep{pintoneto2018}
\begin{equation}
    \frac{\dot{R}^2}{2 {\cal N}^2 R^2} = \frac{\Lambda}{6}-\frac{\kappa}{2 R^2}+Q_{\rm WDW},
    \label{wdw}
\end{equation}
where, for simplicity, we have assumed an empty Universe, and $Q_{\rm WDW}$ is the Wheeler-DeWitt quantum potential. Equation (\ref{wdw}) can be re-cast in the same form as (\ref{f1a}) if we take ${\cal N}(t)=1$, and we set
\begin{equation}
   Q_{\rm WDW} \equiv -\frac{\dot{\lambda}\dot{R}}{\lambda R}-\frac{1}{2}\frac{\dot\lambda^2}{\lambda^2}+\frac{\Lambda (\lambda^2-1)}{6}. 
\end{equation}
However, Equation (\ref{f1a}) is also reproduced if we take ${\cal N}(t)=\lambda(t)$, and make a transformation $R(t) \rightarrow \lambda(t) R(t)$ in (\ref{msp}), \color{black} meaning that quantum effects are equivalent to a coordinate change of the metric, as noted earlier. This also means that, at the level of our approximations, the back-reaction of quantum effects onto the metric is ignored.

Equations (\ref{f1a}) and (\ref{f2a}) are very similar to those of standard cosmology, with some noticeable differences. Firstly, there is now a modified cosmological constant $\tilde{\Lambda}$ which also depends on the quantum potential. Secondly, there is an additional term that produces an acceleration, as with dark energy, but it is not related to the properties of the vacuum. Instead, this new term depends on the non-locality of quantum interactions of a Universe filled with matter. 
%Even in a Universe with $\Lambda=0$ we would have $\tilde{\Lambda}=-3 \dot{\lambda}^2/\lambda^2 = -3 \dot{Q}^2/4 \sim -3 H^2 /R^4 m^4 L^4$, where $H=\dot{R}/R$ is the Hubble parameter and $L$ is the scale of the spatial gradients.
%In practice, quantum effects appear as a negative vacuum energy. In fact, assuming that the scale of density gradients is set by the de-Broglie thermal wavelength of the particles, $L\sim (2\pi/m T)^{1/2}$, we get at present time ($R=1$) $\tilde{\Lambda} \sim -3 H^2 T^2/4 \pi^2 m^2$.

Equation (\ref{f1a}) can be rewritten in a more familiar form as
\begin{equation}
\Omega_m + \Omega_\Lambda+\Omega_\kappa+\Omega_Q=1,
\label{omega}
\end{equation}
where,
\begin{eqnarray}
\Omega_m &=& \frac{8 \pi G \rho}{3 H^2}=\frac{\rho}{\rho_c}, \\
\Omega_\Lambda &=& \frac{\tilde{\Lambda}}{3 H^2}, \\
\Omega_\kappa &=& -\frac{\kappa}{R^2 H^2}, \\
\Omega_Q &=& -\frac{\dot{Q}}{H}-\frac{\dot{Q}^2}{4 H^2},
\end{eqnarray}
with $\rho_c$ the critical density of matter. Observations suggest $\Omega_m \sim 0.3$ at the present time, accounting for both luminous and dark matter \citep{riess1998observational,perlmutter1999measurements,goobar2011supernova,krauss2003age}.

The implication of Eq. (\ref{omega}) in the context of conformal gravity has been discussed in detail in Ref. \cite{maeder2017an,maeder2017dynamical}. Here we consider for simplicity a Universe where, in the weak field limit,
$\tilde{\Lambda} \sim \Lambda = 0$, 
and similarly
the space curvature term is set to zero ($\kappa = 0$).
The latter approximation is consistent with both observational limits \citep{goobar2011supernova} and existing inflationary models \citep{Guth81Inflation}. 
\color{black} Proceeding with this assumption that $\Omega_\Lambda,~\Omega_\kappa \ll 1$, Eq. (\ref{omega}) can be conveniently re-written in the form
\begin{equation}
\Omega_m = \left(1+
\frac{\dot{Q}}{2 H} \right)^2 ,
\label{f3}
\end{equation}
implying that $\dot{Q}<0$, as $\Omega_m<1$. Using our previous definition of $Q$, from Eq.~(\ref{Qdefinition}), the above can be expressed as
\begin{eqnarray}
\frac{d}{dt}\left[\left(\frac{1}{2}-q_{\rm d}\right)H^2\right]
= \frac{4}{3}m^2 H\left(1-\Omega_m^{1/2}\right),
\label{fderiv}
\end{eqnarray}
a non-linear third-order differential equation in $R$. Numerical integration of this equation requires three boundary conditions, at least one of which is uncertain. Nevertheless, we can still comment on the general behaviour. As the right hand side of this equation is clearly positive, consistency requires that 
\begin{equation}
\frac{\dot{q}_{\rm d}}{H} < (q_{\rm d}+1)(2 q_{\rm d}-1) \,
\end{equation}
indicating a number of different interesting regimes: 

\begin{enumerate}
    \item[i)] $q_{\rm d} > 1/2$: in this case, the Universe is decelerating, and consistent solutions for both positive and negative $\dot{q}_{\rm d}$ can be found. Such large deceleration parameters are however disfavoured by current observations.
    \item[ii)] $q_{\rm d} < -1$: i.e. the universe is accelerating rapidly. $\dot{q}_{\rm d}$ again can have both positive and negative values but as previously, this regime is also disfavoured by observations.
    \item[iii)] $-1 \leq q_{\rm d} \leq 1/2$: this is physically the most interesting regime, and can be subdivided further into two sub-cases: one accelerating ($q_{\rm d} <0$), one decelerating ($q_{\rm d} >0$). However, $\dot{q}_{\rm d}<0$ for both implying that a transition from case (iii) to case (i) is not possible. A transition to case (ii) cannot be ruled out. 
    
\end{enumerate}
Current observations favour the latter scenario, with $q_{\rm d} < 0 $ \citep{riess1998observational, perlmutter1999measurements}, although this conclusion is not universally accepted \citep{subir2018}. Proceeding on the assumption that the universe is in fact accelerating, we conclude that the term $\Omega_Q$ behaves as an acceleration that opposes gravity. It is equivalent to what is usually referred to as dark energy \citep{riess1998observational, perlmutter1999measurements}. In this context, as pointed out earlier, it represents the background energy associated to the non-local quantum nature of matter.

%As the Universe is expanding and the density decreasing, the Bohm potential also decreases with time, hence $\dot{Q} \approx -\gamma H^3/m^2<0$. Thus, $\Omega_m<1$ if $|\dot{Q}/4H|=\gamma H^2/4 m^2<1$. If this condition is satisfied, then the term $\Omega_{DE} = -(\dot{Q}/H) (\dot{Q}/4H+1)
%\approx \gamma H^2/m^2$ behaves as an acceleration that opposes gravity. It is equivalent to what is usually referred to as dark energy \cite{riess1998observational,perlmutter1999measurements}. In this context, as pointed out earlier, it represents the background energy associated to the non-local quantum nature of matter.
\color{black}
Using dimensional analysis on equation (\ref{fderiv}),
one can infer the present value of the dark energy density 
$\Omega^0_{DE} = \Omega^0_{Q} \sim  H_0^2/m^2 \sim {\cal O}(1)$, where $H_0 = 70 \,\, {\rm km \,s^{-1}\,Mpc^{-1}}= 2 \times 10^{-33}$ eV \citep{weinberg1989}.
This would imply the existence of a field particle of mass
$m \equiv m_g \sim 2 \times 10^{-33}$ eV. 
We note that this result is, in fact, analogous to what obtained in quintessence (scalar-field) models of dark energy \citep{copeland2006}, whereby the present value of the cosmological constant is explained by a scalar field of mass $m_\phi \equiv m_g \sim H_0$.
The existence of such a field is hypothetical, nevertheless the value given for its mass is consistent with the more stringent bounds on the graviton mass \citep{goldhaberphoton}. There are, however, reservations in this interpretation as it gives the present time a very special place in assigning the mass of such a particle, set only by the current value of the Hubble constant, an example of fine-tuning.

\color{black}
\section{Beyond the isotropic and homogeneous Universe}
\noindent
The model presented so far is still quite ideal. In particular, it is assumed that the Universe is homogeneous and isotropic. However, it is clear from observations that the Universe, mostly at smaller scales, is neither homogeneous nor isotropic. Locally, it consists of gravitationally bound structures, such as cluster of galaxies, and voids. Since the equations of 
general relativity are non-linear, it is plausible that local density inhomogeneities could 
produce significant changes to the cosmological evolution as described by the standard Friedmann's equations. To investigate the effects of an inhomogeneous distribution of mass, let us consider the case of a spherically symmetric dust Universe (as seen from our location at the center) described by the Lema\^itre-Tolman-Bondi metric. To further simplify the analysis, as we did before for the case of the Robertson-Walker metric, we take $\Lambda=0$ and $\kappa=0$.
The line element is thus \citep{enqvist2008lemaitretolmanbondi}:
\begin{widetext}
\begin{eqnarray}
ds^2 = dt^2- \left[\partial_r F(r,t) \right]^2 dr^2 + 
\left[F(r,t) \right]^2 (d\theta^2+\sin^2\theta \, d\phi^2),
\end{eqnarray}
\end{widetext}
where $F(r,t)$ is some (generally unknown) function of the radial coordinate and time. To constrain the model, we now assume that the function $F(r,t)$ can be parametrized in the following form: $F(r,t)=R(t) \left[
r+f(r)\right]$, where $R(t)$ is the same scale factor as in the Robertson-Walker metric and $f(r)$ is a function that describe the departure from homogeneity. This reduces to  
the Robertson-Walker metric (for $\Lambda=0$ and $\kappa=0$) if $f(r)=0$.

As we have done before for the Robertson-Walker metric, in the presence of non-local effects induced by the quantum nature of matter, we also need to include the conformal factor $\lambda^2=e^{Q(r,t)}$, where we have explicitly included both a radial and time dependence in the Bohm potential. With this metric, the solution of Einstein's field equation gives a modified version of Equation (\ref{f1a}):
\begin{widetext}
\begin{eqnarray}
\label{f4}
\scriptsize
\frac{\dot{R}^2}{R^2}&=&\frac{8\pi G}{3}\rho-\frac{(\partial_t Q) \dot{R}}{R} \left[
\frac{(\partial_t Q) R}{4 \dot{R}} + 1\right] +
\frac{\{2+4 f'(r)+2 [f'(r)]^2-[r+f(r)]f''(r)\}(\partial_r Q)}{3 R^2 [r+f(r)] [1+f'(r)]^3} +
\frac{(\partial_r Q)^2+4(\partial_r^2 Q)}{12 R^2 [1+f'(r)]^2}. 
\end{eqnarray}
\end{widetext}
If the conformal factor is independent of the radial coordinate, equation (\ref{f3}) is recovered. However, in an inhomogeneous Universe, the conformal factor will, in general, depend on density and so fluctuations also act toward changing the local expansion rate \citep{buchert2000on,buchert2001on,alnes2006inhomogeneous}, but in a way which is not trivial and depends on the exact structure of those perturbations.

\section{Dark energy in an inhomogeneous Universe}
\noindent
To simplify equation (\ref{f4}) still further, let us take
$Q(r,t)=Q(t) \left[ 1+ \ell_q(r) \right]$ and assume that fluctuations in the metric are uncorrelated with fluctuations of the quantum potential. Such fluctuations are associated to random density perturbations. Locally, however, there are inhomogeneities and anisotropies in the distribution of matter, and a transition to statistical homogeneity is reached only at sufficiently large scales. 
In standard cosmology this scale is taken to be $\sim$100 Mpc \citep{Hogg}, consistent with 
baryon oscillations from the Sloan Digital Sky Survey \citep{2005ApJ_EP}.
On the other hand, there is still the possibility that inhomogeneities of the matter distribution persist at distances exceeding $\sim$300 Mpc \citep{subir2018}, which has then implications on the determination of the Hubble constant \citep{subir2018,Hess}.
Thus when averaging over a volume larger than this, we must have 
$\langle \ell_q(r) \rangle = \langle \partial_r \ell_q(r) \rangle = \langle \partial^2_r \ell_q(r) \rangle = \langle f(r) \rangle = \langle f'(r) \rangle = 0$. Finally, we take the spatial fluctuations to be small in amplitude, that is, $|f'(r)| < 1$ and $|\ell_q(r)| < 1$. \color{black}
Thus, the only non vanishing terms in equation (\ref{f4}) result in
\begin{equation}
\Omega_m  
-\frac{\dot{Q}^2 \langle
\ell_q^2 \rangle}{4 H^2}
+ \frac{Q^2 \langle
(\partial_r \ell_q)^2 \rangle}{12 R^2 H^2}=\left(1+ \frac{\dot{Q}}{2H}  \right)^2.
\end{equation}
Hence, as discussed earlier, using dimensional analysis, $|Q| \approx H^2/m^2$, $|\dot{Q}| \approx H^3/m^2$, and
$|\partial_r \ell_q| \approx \delta/L$, where $L$ is the scale of the fluctuations of the field $m$, and $\delta = \langle \ell_q^2 \rangle^{1/2}$.
Inhomogeneities dominates the total dark energy density only if $L \ll \delta/mR$. If we take this to be the case, the dark energy density has a much simplified relation
\begin{equation}
\Omega_{DE} \approx \frac{\delta^2 H^2}{12 L^2 R^2 m^4}. 
\end{equation}
\color{black}

One possibility we explore now is that this hypothetical field $m$ is due to an ultralight axion. %Axions are a compelling extension to the Standard Model. 
The axion is a pseudo Nambu-Goldstone boson of the broken $U(1)$ Peccei-Quinn symmetry \citep{peccei1977cp,wilczek1978problem}, which was introduced to explain the absence of $CP$ violation in strong  interactions. Extensions of this model to axion-like particles \citep{jaeckel2010the} can admit the existence of ultralight scalars or pseudo-scalars.
The axion remains effectively massless until the Universe cools below some critical temperature, and after that it acquires a mass $m_a$ and starts oscillating with wavelength $\propto 1/m_a$. 
The energy density of these oscillations is of order the critical density of the Universe, hence the ``invisible axion" is a well-motivated candidate for the dark matter \citep{DINE1983137,ABBOTT1983133,PRESKILL1983127}. Because of its small coupling, the axion field decouples with the baryonic matter and behaves as a Bose-Einstein condensate.

The property of axionic dark matter (with $m_a\sim 8 \times 10^{-23}$ eV giving the best fit to observations)  has been investigated in recent structure formation simulations \citep{schive2014cosmic}.
Interestingly, those simulations seem to
explain the mass distribution around dwarf galaxies, which instead cannot be reproduced correctly by structure formation calculations where dark matter is made by cold non-interacting classical particles \citep{schive2014cosmic}. A characteristic of axionic dark matter simulations is a fuzziness of the mass distribution. This is set by the axion Compton wavelength $\lambda_{Ca} = (2\pi/m_a T_a)^{1/2}$, where $T_a=\left[ 2\pi/\zeta(3/2) m_a \right] n_a^{2/3}$ is the critical temperature of the Bose-Einstein condensate \citep{bec}, $\zeta(3/2)=2.61$ is the Riemann zeta function and $n_a = \rho_{DM}/m_a$ is the present-day number density of axions, with $\rho_{DM}=9.6\times 10^{-12}$ eV$^{-4}$ the co-moving dark matter mass density \citep{1674-1137-40-10-100001}.

%Let us assume that at a some time $t_m$ after inflation, the scalar field acquires a mass $m$. We don't know much about the size of spatial fluctuations of such field, but it is plausible these will be sufficiently homogeneous, up to a scale comparable to the Hubble one, that is, $L_m \sim 1/H_m$, where $1/H_m \approx t_m$, and $\delta$ of order of a few percent. The corresponding comoving scale is $L_0 \sim 1/H_m R_m$, where $R_m$ is the scale factor at the time $t_m$. The present-day dark energy is then

We do not know much about the size of the density perturbations. If we assume they are typically of the same order as the fluctuations in temperature seen by the cosmic microwave background, then $\delta \sim 10^{-5}$ \citep{2004mmu..symp..291W}.
Setting $L \sim \lambda_{Ca}$, the equation for the dark matter density at the present epoch becomes
\begin{equation}
\Omega^0_{DE} \approx \frac{\delta^2 H_0^2 \rho^{2/3}_{DM}}{12\,\xi(3/2)\,m_a^{14/3}}. 
\end{equation}
Requiring $\Omega^0_{DE} \approx 0.7$, this gives $m_a \sim 10^{-18}$ eV. 
Such (and even smaller) axion masses are, in fact, what is needed for small scale structure formation \citep{schive2014cosmic} and they may be detected by oscillations in pulsar timings \citep{PhysRevLett.119.221103}. Moreover, small ultralight scalars in the range $10^{-21} \,\,{\rm eV} \lesssim m_a \lesssim 10^{-17}\,\,{\rm eV}$ have been invoked to explain the cosmological origin of magnetic fields \citep{2018arXiv180207269C} (see also Ref.~\cite{PhysRevLett.121.021301}).

\section{Summary and Conclusions}
\noindent
In this work we have argued that the interpretation of the modified Friedmann's equation in conformal gravity including inhomogeneities in the matter is perhaps clarified if the conformal factor is determined by the quantum potential of Bohmian mechanics. In particular, these results would suggest that the acceleration term in the expansion of the Universe need not be set by the vacuum energy but rather by the non-locality of matter. This is by no means a solution of the cosmological constant problem, as the solution is reliant on a term which requires the existence of as yet undiscovered particle fields. 
\vspace{5pt}

%Taking the temperature to be $T=2.3\times 10^{-4}$ eV (2.7 K), and assuming that the major constituent of the Universe is a hypothetical dark
%matter particle of mass $m=0.1$ meV, we get $\Omega_{DE} \sim 0.7$ at present time, in agreement with observations \cite{goobar2011supernova}. 
%While such particle has not been discovered yet, a well motivated candidate is the axion, a pseudo Nambu-Goldstone boson of the broken $U(1)$ Peccei-Quinn symmetry \cite{peccei1977cp,wilczek1978problem} which was introduced to explain the absence of $CP$ violation in strong  interactions. Recent lattice quantum chromodynamics (QCD) calculations \cite{borsanyi2016calculation} have, in fact, found that, in order to explain the present-day dark matter density, the axion mass must be in the range between 50 $\mu$eV to 1.5 meV, which is well compatible with the discussions presented here.

\subsection*{Acknowledgements}
The authors would like to thank Prof Subir Sarkar (University of Oxford) for stimulating discussions related to this work.
This research was supported in part by the Engineering and Physical Sciences Research Council, UK (EP/M022331/1, EP/N014472/1, EP/N002644/1, EP/P010059/1).

\subsection*{Author contributions}
This project was conceived by G.G.
Calculations were performed by B.R. and G.G., with B.L. contributing to the extension to many-body systems. The paper was written by all the authors. 

\subsection*{Data availability}
All data that support the findings of this study are available from
the authors upon request.

%\subsection*{Author declaration}
%The authors declare no competing interests as defined by Nature Research, or other interests that might be perceived to influence the results and/or discussion reported in this paper.

%


\begin{thebibliography}{59}%
\makeatletter
\providecommand \@ifxundefined [1]{%
 \@ifx{#1\undefined}
}%
\providecommand \@ifnum [1]{%
 \ifnum #1\expandafter \@firstoftwo
 \else \expandafter \@secondoftwo
 \fi
}%
\providecommand \@ifx [1]{%
 \ifx #1\expandafter \@firstoftwo
 \else \expandafter \@secondoftwo
 \fi
}%
\providecommand \natexlab [1]{#1}%
\providecommand \enquote  [1]{``#1''}%
\providecommand \bibnamefont  [1]{#1}%
\providecommand \bibfnamefont [1]{#1}%
\providecommand \citenamefont [1]{#1}%
\providecommand \href@noop [0]{\@secondoftwo}%
\providecommand \href [0]{\begingroup \@sanitize@url \@href}%
\providecommand \@href[1]{\@@startlink{#1}\@@href}%
\providecommand \@@href[1]{\endgroup#1\@@endlink}%
\providecommand \@sanitize@url [0]{\catcode `\\12\catcode `\$12\catcode
  `\&12\catcode `\#12\catcode `\^12\catcode `\_12\catcode `\%12\relax}%
\providecommand \@@startlink[1]{}%
\providecommand \@@endlink[0]{}%
\providecommand \url  [0]{\begingroup\@sanitize@url \@url }%
\providecommand \@url [1]{\endgroup\@href {#1}{\urlprefix }}%
\providecommand \urlprefix  [0]{URL }%
\providecommand \Eprint [0]{\href }%
\providecommand \doibase [0]{http://dx.doi.org/}%
\providecommand \selectlanguage [0]{\@gobble}%
\providecommand \bibinfo  [0]{\@secondoftwo}%
\providecommand \bibfield  [0]{\@secondoftwo}%
\providecommand \translation [1]{[#1]}%
\providecommand \BibitemOpen [0]{}%
\providecommand \bibitemStop [0]{}%
\providecommand \bibitemNoStop [0]{.\EOS\space}%
\providecommand \EOS [0]{\spacefactor3000\relax}%
\providecommand \BibitemShut  [1]{\csname bibitem#1\endcsname}%
\let\auto@bib@innerbib\@empty
%</preamble>
\bibitem [{\citenamefont {Hiley}\ and\ \citenamefont
  {Muft}(1995)}]{hiley1995the}%
  \BibitemOpen
  \bibfield  {author} {\bibinfo {author} {\bibfnamefont {Hiley},\ \bibnamefont
  {B.J.}}\ \&\ \bibinfo {author} {\bibfnamefont {Muft},\ \bibnamefont
  {A.A.}},\ }\href {\doibase 10.1007/978-94-015-8529-3_15} {\emph {\bibinfo
  {title} {The ontological interpretation of quantum field theory applied in a
  cosmological context}}}\ (\bibinfo  {publisher} {Springer},\ \bibinfo {year}
  {1995})\ pp.\ \bibinfo {pages} {141--156}\BibitemShut {}%
\bibitem [{\citenamefont {Bohm}(1952)}]{bohm1952a}%
  \BibitemOpen
  \bibfield  {author} {\bibinfo {author} {\bibfnamefont {Bohm},~\bibnamefont
  {D.}},\ }\href@noop {} {\emph {\bibinfo
  {title} {A Suggested Interpretation of the Quantum Theory of ``Hidden" Variables. I}}}, {\bibfield  {journal} {\bibinfo  {journal} {Physical
  Review}\ }\textbf {\bibinfo {volume} {85}},\ \bibinfo {pages} {166} (\bibinfo
  {year} {1952})}\BibitemShut {}%
\bibitem [{\citenamefont {Bohm}\ \emph {et~al.}(1987)\citenamefont {Bohm},
  \citenamefont {Hiley},\ and\ \citenamefont {Kaloyerou}}]{bohm1987an}%
  \BibitemOpen
  \bibfield  {author} {\bibinfo {author} {\bibfnamefont {Bohm},~\bibnamefont
  {D.}}, \bibinfo {author} {\bibfnamefont {Hiley}~\bibnamefont {B.}}, \ \&\
  \bibinfo {author} {\bibfnamefont {Kaloyerou}~\bibnamefont {P.}},\ } {\emph {\bibinfo
  {title} {An ontological basis for the quantum theory}}},  \href
  {\doibase 10.1016/0370-1573(87)90024-x} {\bibfield  {journal} {\bibinfo
  {journal} {Physics Reports}\ }\textbf {\bibinfo {volume} {144}},\ \bibinfo
  {pages} {321} (\bibinfo {year} {1987})}\BibitemShut {}%
\bibitem [{\citenamefont {Mahler}\ \emph {et~al.}(2016)\citenamefont {Mahler},
  \citenamefont {Rozema}, \citenamefont {Fisher}, \citenamefont {Vermeyden},
  \citenamefont {Resch}, \citenamefont {Wiseman},\ and\ \citenamefont
  {Steinberg}}]{mahler2016experimental}%
  \BibitemOpen
  \bibfield  {author} {\bibinfo {author} {\bibfnamefont {Mahler},\ \bibnamefont
  {D.H.}}, \bibinfo {author} {\bibfnamefont {Rozema},~\bibnamefont {L.}},
  \bibinfo {author} {\bibfnamefont {Fisher},~\bibnamefont {K.}}, \bibinfo
  {author} {\bibfnamefont {Vermeyden},~\bibnamefont {L.}}, \bibinfo {author}
  {\bibfnamefont {Resch},\ \bibnamefont {K.J.}}, \bibinfo {author}
  {\bibfnamefont {Wiseman},\ \bibnamefont {H.M.}}, \ \&\ \bibinfo {author}
  {\bibfnamefont {Steinberg},~\bibnamefont {A.}},\ } {\emph {\bibinfo
  {title} {Experimental nonlocal and surreal Bohmian trajectories}}}, \href {\doibase
  10.1126/sciadv.1501466} {\bibfield  {journal} {\bibinfo  {journal} {Science
  Advances}\ }\textbf {\bibinfo {volume} {2}},\ \bibinfo {pages} {e1501466}
  (\bibinfo {year} {2016})}\BibitemShut {}%
\bibitem [{\citenamefont {{Haas}}(2011)}]{HaasQuantumPlasmas}%
  \BibitemOpen
  \bibfield  {author} {\bibinfo {author} {\bibfnamefont {Haas},~\bibnamefont
  {{F.}}},\ }\href {\doibase 10.1007/978-1-4419-8201-8} {\emph {\bibinfo
  {title} {Quantum Plasmas: An Hydrodynamic Approach}}}\ (\bibinfo {year}
  {Springer, 2011})\BibitemShut {}%
\bibitem [{\citenamefont {Cross}\ \emph {et~al.}(2014)\citenamefont {Cross},
  \citenamefont {Reville},\ and\ \citenamefont {Gregori}}]{cross}%
  \BibitemOpen
  \bibfield  {author} {\bibinfo {author} {\bibfnamefont {Cross},\ \bibnamefont
  {J.E.}}, \bibinfo {author} {\bibfnamefont {Reville},~\bibnamefont {B.}}, \
  \&\ \bibinfo {author} {\bibfnamefont {Gregori},~\bibnamefont {G.}},\ } {\emph {\bibinfo
  {title} {Scaling of magneto-quantum-radiative hydrodynamic equations: from laser-produced plasmas to astrophysics}}}, \href
  {http://stacks.iop.org/0004-637X/795/i=1/a=59} {\bibfield  {journal}
  {\bibinfo  {journal} {The Astrophysical Journal}\ }\textbf {\bibinfo {volume}
  {795}},\ \bibinfo {pages} {59} (\bibinfo {year} {2014})}\BibitemShut
  {}%
\bibitem [{\citenamefont {D\"urr}\ \emph {et~al.}(2004)\citenamefont {D\"urr},
  \citenamefont {Goldstein}, \citenamefont {Tumulka},\ and\ \citenamefont
  {Zanghì}}]{drr2004bohmian}%
  \BibitemOpen
  \bibfield  {author} {\bibinfo {author} {\bibfnamefont {D\"urr},~\bibnamefont
  {D.}}, \bibinfo {author} {\bibfnamefont {Goldstein},~\bibnamefont {S.}},
  \bibinfo {author} {\bibfnamefont {Tumulka},~\bibnamefont {R.}}, \ \&\
  \bibinfo {author} {\bibfnamefont {Zangh\`i},~\bibnamefont {N.}},\ }
   {\emph {\bibinfo
  {title} {Bohmian Mechanics and Quantum Field Theory}}}, \href
  {\doibase 10.1103/physrevlett.93.090402} {\bibfield  {journal} {\bibinfo
  {journal} {Physical Review Letters}\ }\textbf {\bibinfo {volume} {93}},\
  \bibinfo {pages} {090402} (\bibinfo {year} {2004})}\BibitemShut {}%
\bibitem [{\citenamefont {Carroll}(2007)}]{carroll2007metric}%
  \BibitemOpen
  \bibfield  {author} {\bibinfo {author} {\bibfnamefont {Carroll},~\bibnamefont
  {R.}},\ }  {\emph {\bibinfo
  {title} {Metric fluctuations, entropy, and the Wheeler-deWitt equation}}}, \href {\doibase 10.1007/s11232-007-0076-2} {\bibfield
  {journal} {\bibinfo  {journal} {Theoretical and Mathematical Physics}\
  }\textbf {\bibinfo {volume} {152}},\ \bibinfo {pages} {904} (\bibinfo {year}
  {2007})}\BibitemShut {}%
\bibitem [{\citenamefont {D\"urr}\ \emph {et~al.}(2014)\citenamefont {D\"urr},
  \citenamefont {Goldstein}, \citenamefont {Norsen}, \citenamefont {Struyve},\
  and\ \citenamefont {Zanghì}}]{drr2014can}%
  \BibitemOpen
  \bibfield  {author} {\bibinfo {author} {\bibfnamefont {D\"urr},~\bibnamefont
  {D.}}, \bibinfo {author} {\bibfnamefont {Goldstein},~\bibnamefont {S.}},
  \bibinfo {author} {\bibfnamefont {Norsen},~\bibnamefont {T.}}, \bibinfo
  {author} {\bibfnamefont {Struyve},~\bibnamefont {W.}}, \ \&\ \bibinfo
  {author} {\bibfnamefont {Zangh\`i},~\bibnamefont {N.}},\ }
  {\emph {\bibinfo
  {title} {Can Bohmian mechanics be made relativistic?}}} \href {\doibase
  10.1098/rspa.2013.0699} {\bibfield  {journal} {\bibinfo  {journal} {Proc. R.
  Soc. A}\ }\textbf {\bibinfo {volume} {470}},\ \bibinfo {pages} {20130699}
  (\bibinfo {year} {2014})}\BibitemShut {}%
\bibitem [{\citenamefont {Holland}(1992)}]{holland1992the}%
  \BibitemOpen
  \bibfield  {author} {\bibinfo {author} {\bibfnamefont {Holland},\ \bibnamefont
  {P.R.}},\ } {\emph {\bibinfo
  {title} {The Dirac equation in the de Broglie-Bohm theory of motion}}}, \href {\doibase 10.1007/bf01889714} {\bibfield  {journal}
  {\bibinfo  {journal} {Foundations of Physics}\ }\textbf {\bibinfo {volume}
  {22}},\ \bibinfo {pages} {1287} (\bibinfo {year} {1992})}\BibitemShut
  {}%
\bibitem [{\citenamefont {Shojai}\ and\ \citenamefont
  {Shojai}(2006)}]{shojai2006about}%
  \BibitemOpen
  \bibfield  {author} {\bibinfo {author} {\bibfnamefont {Shojai},~\bibnamefont
  {A.}}\ \&\ \bibinfo {author} {\bibfnamefont {Shojai},~\bibnamefont
  {F.}},\ }  {\emph {\bibinfo
  {title} {About Some Problems Raised by the Relativistic Form of De-Broglie–Bohm Theory of Pilot Wave}}}, \href {\doibase 10.1238/Physica.Regular.064a00413} {\bibfield
  {journal} {\bibinfo  {journal} {Physica Scripta}\ }\textbf {\bibinfo {volume}
  {64}},\ \bibinfo {pages} {413} (\bibinfo {year} {2006})}\BibitemShut
  {}%
\bibitem [{\citenamefont {Carroll}(2005)}]{carroll2005fluctuations}%
  \BibitemOpen
  \bibfield  {author} {\bibinfo {author} {\bibfnamefont {Carroll},~\bibnamefont
  {R.}},\ }  {\emph {\bibinfo
  {title} {Fluctuations, gravity, and the quantum potential}}}, \href {http://arxiv.org/abs/gr-qc/0501045} {\bibfield
  {journal} {\bibinfo  {journal} {arXiv preprint gr-qc/0501045}\ } (\bibinfo
  {year} {2005})}\BibitemShut {}%
  
\bibitem [{\citenamefont
  {Goldstein}(1950)}]{goldstein1950}%
  \BibitemOpen
  \bibfield  {author} {\bibinfo {author} {\bibfnamefont {Goldstein},\ \bibnamefont
  {H.}},\ } {\emph {\bibinfo
  {title} {Classical Mechanics}}}\ (\bibinfo  {publisher} {Addison-Wesley},\ \bibinfo {year}
  {1950})\BibitemShut {}%  
\bibitem [{\citenamefont
  {Zee}(2013)}]{zee2013}%
  \BibitemOpen
  \bibfield  {author} {\bibinfo {author} {\bibfnamefont {Zee},\ \bibnamefont
  {A.}},\ } {\emph {\bibinfo
  {title} {Einstein Gravity in a Nutshell}}}\ (\bibinfo  {publisher} {Princeton},\ \bibinfo {year}
  {2013})\BibitemShut {}%
\bibitem [{\citenamefont {Durr}\ \citenamefont
  {Zee}(2013)}]{durr}%
  \BibitemOpen
  \bibfield  {author} {\bibinfo {author} {\bibfnamefont {D\"urr},\ \bibnamefont
  {D.}}\ \&\ \bibinfo {author} {\bibfnamefont {Teufel},\ \bibnamefont
  {S.}},\ } {\emph {\bibinfo
  {title} {Bohmian Mechanics}}}\ (\bibinfo  {publisher} {Springer},\ \bibinfo {year}
  {2009})\BibitemShut {}%  
\bibitem [{\citenamefont {Manfredi}\ and\ \citenamefont
  {Haas}(2001)}]{manfredi2001selfconsistent}%
  \BibitemOpen
  \bibfield  {author} {\bibinfo {author} {\bibfnamefont {Manfredi},~\bibnamefont
  {G.}}\ \&\ \bibinfo {author} {\bibfnamefont {Haas},~\bibnamefont
  {F.}},\ }  {\emph {\bibinfo
  {title} {Self-consistent fluid model for a quantum electron gas}}}, \href {\doibase 10.1103/physrevb.64.075316} {\bibfield  {journal}
  {\bibinfo  {journal} {Physical Review B}\ }\textbf {\bibinfo {volume} {64}},\
  \bibinfo {pages} {075316} (\bibinfo {year} {2001})}\BibitemShut {}%
\bibitem [{\citenamefont {Mitcha}\ \emph {et~al.}(2015)}]{graziani}%
  \BibitemOpen
  \bibfield  {author} {\bibinfo {author} {\bibfnamefont {Mitcha},~\bibnamefont
  {D.}},\ \bibinfo {author} {\bibfnamefont {Graziani},~\bibnamefont
  {F.}}\ \&\ \bibinfo {author} {\bibfnamefont {Bonitz},~\bibnamefont
  {M.}},\ } {\emph {\bibinfo
  {title} {Quantum Hydrodynamics for Plasmas -- a Thomas‐Fermi Theory Perspective}}}, \href {\doibase 10.1002/ctpp.201500024} {\bibfield  {journal}
  {\bibinfo  {journal} {Contributions in Plasma Physics}\ }\textbf {\bibinfo {volume} {55}},\
  \bibinfo {pages} {437--443} (\bibinfo {year} {2015})}\BibitemShut {}%
\bibitem [{\citenamefont {Moldabekov}\ \emph {et~al.}(2015)}]{Moldabekov}%
  \BibitemOpen
  \bibfield  {author} {\bibinfo {author} {\bibfnamefont {Moldabekov},~\bibnamefont
  {Zh. A.}},\ \bibinfo {author} {\bibfnamefont {Bonitz},~\bibnamefont
  {M.}}\ \&\ \bibinfo {author} {\bibfnamefont {Ramazanov},~\bibnamefont
  {T. S.}},\ } {\emph {\bibinfo
  {title} {Theoretical foundations of quantum hydrodynamics for plasmas}}}, \href {\doibase 10.1063/1.5003910} {\bibfield  {journal}
  {\bibinfo  {journal} {Phys. Plasmas}\ }\textbf {\bibinfo {volume} {25}},\
  \bibinfo {pages} {031903} (\bibinfo {year} {2018})}\BibitemShut {}%
  
\bibitem[{\citenamefont {Hohenberg \& Kohn}\ (1964)}]{Hohenberg1964}
Hohenberg, P. \& Kohn, W.,
\newblock{\em Inhomogeneous electron gas},
\newblock{Physical Review} {\bf 136}, B864 (1964).

\bibitem[{\citenamefont {Kohn \& Sham}\ (1965)}]{Kohn1965}
Kohn, W. \& Sham, L.~J.,
\newblock{\em Self-consistent equations
           including exchange and correlation effects},
\newblock{Physical Review} {\bf 140}, A1133 (1965).

\bibitem[Larder et al.(2019)]{2018arXiv181108161L} Larder, B., Gericke, D., Richardson, S., et al.\, 
{\em Fast Non-Adiabatic Dynamics of Many-Body Quantum Systems},
{Science Advances} (accepted for publication), arXiv:1811.08161 (2019). 


\bibitem [{\citenamefont {Nikoli{\'c}}(2005)}]{nikoli2005relativistic}%
  \BibitemOpen
  \bibfield  {author} {\bibinfo {author} {\bibfnamefont {Nikoli\'c},~\bibnamefont
  {H.}},\ } {\emph {\bibinfo
  {title} {Relativistic Quantum Mechanics and the Bohmian Interpretation}}}, \href {\doibase 10.1007/s10702-005-1128-1} {\bibfield
  {journal} {\bibinfo  {journal} {Foundations of Physics Letters}\ }\textbf
  {\bibinfo {volume} {18}},\ \bibinfo {pages} {549} (\bibinfo {year}
  {2005})}\BibitemShut {}%
\bibitem [{\citenamefont {Nikoli{\'c}}(2007)}]{nikoli2007bohmian}%
  \BibitemOpen
  \bibfield  {author} {\bibinfo {author} {\bibfnamefont {Nikoli\'c},~\bibnamefont
  {H.}},\ } {\emph {\bibinfo
  {title} {Bohmian mechanics in relativistic quantum mechanics, quantum field theory and string theory}}}, \href {\doibase 10.1088/1742-6596/67/1/012035} {\bibfield
  {journal} {\bibinfo  {journal} {Journal of Physics: Conference Series}\
  }\textbf {\bibinfo {volume} {67}},\ \bibinfo {pages} {012035} (\bibinfo
  {year} {2007})}\BibitemShut {}%
\bibitem [{\citenamefont {Gonz\'alez-D\'iaz}(2004)}]{gonzlezdaz2004subquantum}%
  \BibitemOpen
  \bibfield  {author} {\bibinfo {author} {\bibfnamefont {Gonz\'alez-D\'iaz},\ \bibnamefont
  {P.F.}},\ } {\emph {\bibinfo
  {title} {Subquantum dark energy}}}, \href {\doibase 10.1103/PhysRevD.69.103512} {\bibfield
  {journal} {\bibinfo  {journal} {Physical Review D}\ }\textbf {\bibinfo
  {volume} {69}},\ \bibinfo {pages} {103512} (\bibinfo {year}
  {2004})}\BibitemShut {}%
  
  \bibitem [{\citenamefont
  {de Broglie}(1960)}]{deBroglie}%
  \BibitemOpen
  \bibfield  {author} {\bibinfo {author} {\bibfnamefont {de Broglie},\ \bibnamefont
  {L.}},\ } {\emph {\bibinfo
  {title} {Nonlinear wave mechanics: A causal interpretation}}}\ (\bibinfo  {publisher} {Elsevier, Amsterdam},\ \bibinfo {year}
  {1960})\BibitemShut {}%  

  
\bibitem [{\citenamefont {Das}(2014)}]{das2014quantum}%
  \BibitemOpen
  \bibfield  {author} {\bibinfo {author} {\bibfnamefont {Das},~\bibnamefont
  {S.}},\ }  {\emph {\bibinfo
  {title} {Quantum Raychaudhuri equation}}}, \href {\doibase 10.1103/PhysRevD.89.084068} {\bibfield  {journal}
  {\bibinfo  {journal} {Physical Review D}\ }\textbf {\bibinfo {volume} {89}},\
  \bibinfo {pages} {084068} (\bibinfo {year} {2014})}\BibitemShut {}%

\bibitem [{\citenamefont {Ali}\ and\ \citenamefont
  {Das}(2015)}]{ali2015cosmology}%
  \BibitemOpen
  \bibfield  {author} {\bibinfo {author} {\bibfnamefont {Ali},~\bibnamefont
  {A.}}\ \&\ \bibinfo {author} {\bibfnamefont {Das},~\bibnamefont {S.}},\
  }  {\emph {\bibinfo
  {title} {Cosmology from quantum potential}}}, \href {\doibase 10.1016/j.physletb.2014.12.057} {\bibfield  {journal}
  {\bibinfo  {journal} {Physics Letters B}\ }\textbf {\bibinfo {volume}
  {741}},\ \bibinfo {pages} {276} (\bibinfo {year} {2015})}\BibitemShut
  {}%
\bibitem [{\citenamefont {Canuto}\ \emph {et~al.}(1977)\citenamefont {Canuto},
  \citenamefont {Hsieh},\ and\ \citenamefont
  {Adams}}]{canuto1977scalecovariant}%
  \BibitemOpen
  \bibfield  {author} {\bibinfo {author} {\bibfnamefont {Canuto},~\bibnamefont
  {V.}}, \bibinfo {author} {\bibfnamefont {Hsieh},\ \bibnamefont {S.H.}}, \
  \&\ \bibinfo {author} {\bibfnamefont {Adams},\ \bibnamefont {P.J.}},\ } {\emph {\bibinfo
  {title} {Scale-Covariant Theory of Gravitation and Astrophysical Applications}}}, \href
  {\doibase 10.1103/physrevlett.39.429} {\bibfield  {journal} {\bibinfo
  {journal} {Physical Review Letters}\ }\textbf {\bibinfo {volume} {39}},\
  \bibinfo {pages} {429} (\bibinfo {year} {1977})}\BibitemShut {}%
\bibitem [{\citenamefont {Maeder}(2017{\natexlab{a}})}]{maeder2017an}%
  \BibitemOpen
  \bibfield  {author} {\bibinfo {author} {\bibfnamefont {Maeder},~\bibnamefont
  {A.}},\ } {\emph {\bibinfo
  {title} {An alternative to the $\Lambda$CDM model: the case of scale invariance}}}, \href {\doibase 10.3847/1538-4357/834/2/194} {\bibfield
  {journal} {\bibinfo  {journal} {The Astrophysical Journal}\ }\textbf
  {\bibinfo {volume} {834}},\ \bibinfo {pages} {194} (\bibinfo {year}
  {2017}{\natexlab{a}})}\BibitemShut {}%
\bibitem [{\citenamefont {Canuto}\ and\ \citenamefont
  {Lee}(1977)}]{canuto1977the}%
  \BibitemOpen
  \bibfield  {author} {\bibinfo {author} {\bibfnamefont {Canuto},~\bibnamefont
  {V.}},\ \&\ \bibinfo {author} {\bibfnamefont {Lee},~\bibnamefont {J.}},\
  } {\emph {\bibinfo
  {title} {The cosmological constant, broken gauge theories and 3 K black-body radiation}}}, \href {\doibase 10.1016/0370-2693(77)90722-5} {\bibfield  {journal}
  {\bibinfo  {journal} {Physics Letters B}\ }\textbf {\bibinfo {volume} {72}},\
  \bibinfo {pages} {281} (\bibinfo {year} {1977})}\BibitemShut {}%
 \bibitem [{\citenamefont {Pinto-Neto and Struyve}(2018)}]{pintoneto2018}%
  \BibitemOpen
  \bibfield  {author} {\bibinfo {author} {\bibfnamefont {Pinto-Neto},~\bibnamefont
  {N.}},\ \&\ \bibinfo {author} {\bibfnamefont {Struyve},~\bibnamefont {W.}},\ } {\emph {\bibinfo
  {title} {Bohmian quantum gravity and cosmology}}}, \href {https://arxiv.org/abs/1801.03353} {\bibfield  {journal}
  {\bibinfo  {journal} {arXiv:1801.03353}\ }(\bibinfo {year} {2018})}\BibitemShut {}%  
\bibitem [{\citenamefont {Weinberg}(1989)}]{weinberg1989}%
  \BibitemOpen
  \bibfield  {author} {\bibinfo {author} {\bibfnamefont {Weinberg},~\bibnamefont
  {S.}},\ } {\emph {\bibinfo
  {title} {The cosmological constant problem}}}, \href {\doibase 10.1103/RevModPhys.61.1} {\bibfield  {journal}
  {\bibinfo  {journal} {Reviews of Modern Physics}\ }\textbf {\bibinfo {volume}
  {61}},\ \bibinfo {pages} {1} (\bibinfo {year} {1989})}\BibitemShut {}%
\bibitem [{\citenamefont {Sahni}\ and\ \citenamefont
  {Krasiński}(2008)}]{sahni2008republication}%
  \BibitemOpen
  \bibfield  {author} {\bibinfo {author} {\bibfnamefont {Sahni},~\bibnamefont
  {V.}}\ \&\ \bibinfo {author} {\bibfnamefont {Krasi\'nski},~\bibnamefont
  {A.}},\ } {\emph {\bibinfo
  {title} {Republication of: The cosmological constant and the theory of elementary particles (By Ya.B. Zeldovich)}}}, \href {\doibase 10.1007/s10714-008-0624-6} {\bibfield
  {journal} {\bibinfo  {journal} {General Relativity and Gravitation}\ }\textbf
  {\bibinfo {volume} {40}},\ \bibinfo {pages} {1557} (\bibinfo {year}
  {2008})}\BibitemShut {}%
\bibitem [{\citenamefont {Coleman}(1988)}]{coleman1988why}%
  \BibitemOpen
  \bibfield  {author} {\bibinfo {author} {\bibfnamefont {Coleman},~\bibnamefont
  {S.}},\ } {\emph {\bibinfo
  {title} {Why there is nothing rather than something: A theory of the cosmological constant}}}, \href {\doibase 10.1016/0550-3213(88)90097-1} {\bibfield
  {journal} {\bibinfo  {journal} {Nuclear Physics B}\ }\textbf {\bibinfo
  {volume} {310}},\ \bibinfo {pages} {643} (\bibinfo {year}
  {1988})}\BibitemShut {}%
\bibitem [{\citenamefont {Hawking}(1984)}]{hawking1984the}%
  \BibitemOpen
  \bibfield  {author} {\bibinfo {author} {\bibfnamefont {Hawking},~\bibnamefont
  {S.}},\ } {\emph {\bibinfo
  {title} {The cosmological constant is probably zero}}}, \href {\doibase 10.1016/0370-2693(84)91370-4} {\bibfield
  {journal} {\bibinfo  {journal} {Physics Letters B}\ }\textbf {\bibinfo
  {volume} {134}},\ \bibinfo {pages} {403} (\bibinfo {year}
  {1984})}\BibitemShut {}%
\bibitem [{\citenamefont {{Volovik}}(2006)}]{volovik2006from}%
  \BibitemOpen
  \bibfield  {author} {\bibinfo {author} {\bibfnamefont {Volovik},\ \bibnamefont
  {{G.E.}}},\ } {\emph {\bibinfo
  {title} {From Quantum Hydrodynamics to Quantum Gravity}}}, \href@noop {} {\bibfield  {journal} {\bibinfo  {journal}
  }} \Eprint {http://arxiv.org/abs/gr-qc/0612134} {arXiv:gr-qc/0612134 (2006)}
  \BibitemShut {}%
\bibitem [{\citenamefont {Bucher}\ and\ \citenamefont
  {Spergel}(1999)}]{bucher1999is}%
  \BibitemOpen
  \bibfield  {author} {\bibinfo {author} {\bibfnamefont {Bucher},~\bibnamefont
  {M.}},\ \&\ \bibinfo {author} {\bibfnamefont {Spergel},\ \bibnamefont
  {D.M.}},\ } {\emph {\bibinfo
  {title} {Is the dark matter a solid?}}} \href
  {https://journals.aps.org/prd/abstract/10.1103/PhysRevD.60.043505} {\bibfield
  {journal} {\bibinfo  {journal} {Phys. Rev. D}\ }\textbf {\bibinfo {volume}
  {60}},\ \bibinfo {pages} {043505} (\bibinfo {year} {1999})}\BibitemShut
  {}%
\bibitem [{\citenamefont {Volovik}(2001)}]{volovik2001superfluid}%
  \BibitemOpen
  \bibfield  {author} {\bibinfo {author} {\bibfnamefont {Volovik},~\bibnamefont
  {G.E.}},\ } {\emph {\bibinfo
  {title} {Superfluid analogies of cosmological phenomena}}}, \href {\doibase 10.1016/s0370-1573(00)00139-3} {\bibfield
  {journal} {\bibinfo  {journal} {Physics Reports}\ }\textbf {\bibinfo {volume}
  {351}},\ \bibinfo {pages} {195} (\bibinfo {year} {2001})}\BibitemShut
  {}%
\bibitem [{\citenamefont {Barrow}\ and\ \citenamefont
  {Shaw}(2011)}]{barrow2011new}%
  \BibitemOpen
  \bibfield  {author} {\bibinfo {author} {\bibfnamefont {Barrow},\ \bibnamefont
  {J.D.}},\ \&\ \bibinfo {author} {\bibfnamefont {Shaw},\ \bibnamefont
  {D.J.}},\ } {\emph {\bibinfo
  {title} {New Solution of the Cosmological Constant Problems}}}, \href {\doibase 10.1103/physrevlett.106.101302} {\bibfield
  {journal} {\bibinfo  {journal} {Physical Review Letters}\ }\textbf {\bibinfo
  {volume} {106}},\ \bibinfo {pages} {101302} (\bibinfo {year}
  {2011})}\BibitemShut {}%
\bibitem [{\citenamefont {Ng}\ and\ \citenamefont {van
  Dam}(1990)}]{ng1990possible}%
  \BibitemOpen
  \bibfield  {author} {\bibinfo {author} {\bibfnamefont {Ng},\ \bibnamefont
  {Y.J.}},\ \&\ \bibinfo {author} {\bibfnamefont {van Dam}},~\bibnamefont {H.},\
  } {\emph {\bibinfo
  {title} {Possible solution to the cosmological-constant problem}}}, \href {\doibase 10.1103/PhysRevLett.65.1972} {\bibfield  {journal} {\bibinfo
   {journal} {Phys. Rev. Lett.}\ }\textbf {\bibinfo {volume} {65}},\ \bibinfo
  {pages} {1972} (\bibinfo {year} {1990})}\BibitemShut {}%
\bibitem [{\citenamefont {Vink}(1992)}]{vink1992}%
  \BibitemOpen
  \bibfield  {author} {\bibinfo {author} {\bibfnamefont {Vink},\ \bibnamefont
  {J.C.}},\ } {\emph {\bibinfo
  {title} {Quantum potential interpretation of the wave function of the universe}}}, {\bibfield  {journal} {\bibinfo
   {journal} {Phys. Lett. B}\ }\textbf {\bibinfo {volume} {369}},\ \bibinfo
  {pages} {707--728} (\bibinfo {year} {1992})}\BibitemShut {}%
\bibitem [{\citenamefont {Riess}\ \emph {et~al.}(1998)\citenamefont {Riess},
  \citenamefont {Filippenko}, \citenamefont {Challis}, \citenamefont
  {Clocchiatti}, \citenamefont {Diercks}, \citenamefont {Garnavich},
  \citenamefont {Gilliland}, \citenamefont {Hogan}, \citenamefont {Jha},
  \citenamefont {Kirshner}, \citenamefont {Leibundgut}, \citenamefont
  {Phillips}, \citenamefont {Reiss}, \citenamefont {Schmidt}, \citenamefont
  {Schommer}, \citenamefont {Smith}, \citenamefont {Spyromilio}, \citenamefont
  {Stubbs}, \citenamefont {Suntzeff},\ and\ \citenamefont
  {Tonry}}]{riess1998observational}%
  \BibitemOpen
  \bibfield  {author} {\bibinfo {author} {\bibfnamefont {Riess},\ \bibnamefont
  {A.G.}}, \bibinfo {author} {\bibfnamefont {Filippenko},\ \bibnamefont
  {A.V.}}, \bibinfo {author} {\bibfnamefont {Challis},~\bibnamefont {P.}},
  \bibinfo {author} {\bibfnamefont {Clocchiatti},~\bibnamefont {A.}}, \bibinfo
  {author} {\bibfnamefont {Diercks},~\bibnamefont {A.}}, \bibinfo {author}
  {\bibfnamefont {Garnavich},\ \bibnamefont {P.M.}}, \bibinfo {author}
  {\bibfnamefont {Gilliland},\ \bibnamefont {R.L.}}, \bibinfo {author}
  {\bibfnamefont {Hogan},\ \bibnamefont {C.J.}}, \bibinfo {author}
  {\bibfnamefont {Jha},~\bibnamefont {S.}}, \bibinfo {author} {\bibfnamefont
  {Kirshner},\ \bibnamefont {R.P.}}, \bibinfo {author} {\bibfnamefont
  {Leibundgut},~\bibnamefont {B.}}, \bibinfo {author} {\bibfnamefont {Phillips},\
  \bibnamefont {M.M.}}, \bibinfo {author} {\bibfnamefont {Reiss},~\bibnamefont
  {D.}}, \bibinfo {author} {\bibfnamefont {Schmidt},\ \bibnamefont {B.P.}},
  \bibinfo {author} {\bibfnamefont {Schommer},\ \bibnamefont {R.A.}}, \bibinfo
  {author} {\bibfnamefont {Smith},\ \bibnamefont {R.C.}}, \bibinfo {author}
  {\bibfnamefont {Spyromilio},~\bibnamefont {J.}}, \bibinfo {author}
  {\bibfnamefont {Stubbs},~\bibnamefont {C.}}, \bibinfo {author} {\bibfnamefont
  {Suntzeff},\ \bibnamefont {N.B.}}, \ \&\ \bibinfo {author} {\bibfnamefont
  {Tonry},~\bibnamefont {J.}},\ } {\emph {\bibinfo
  {title} {Observational Evidence from Supernovae for an Accelerating Universe and a Cosmological Constant}}}, \href {\doibase 10.1086/300499} {\bibfield
  {journal} {\bibinfo  {journal} {The Astronomical Journal}\ }\textbf {\bibinfo
  {volume} {116}},\ \bibinfo {pages} {1009} (\bibinfo {year}
  {1998})}\BibitemShut {}%
\bibitem [{\citenamefont {Perlmutter}\ \emph {et~al.}(1999)\citenamefont
  {Perlmutter}, \citenamefont {Aldering}, \citenamefont {Goldhaber},
  \citenamefont {Knop}, \citenamefont {Nugent}, \citenamefont {Castro},
  \citenamefont {Deustua}, \citenamefont {Fabbro}, \citenamefont {Goobar},
  \citenamefont {Groom}, \citenamefont {Hook}, \citenamefont {Kim},
  \citenamefont {Kim}, \citenamefont {Lee}, \citenamefont {Nunes},
  \citenamefont {Pain}, \citenamefont {Pennypacker}, \citenamefont {Quimby},
  \citenamefont {Lidman}, \citenamefont {Ellis}, \citenamefont {Irwin},
  \citenamefont {McMahon}, \citenamefont {Ruiz‐Lapuente}, \citenamefont
  {Walton}, \citenamefont {Schaefer}, \citenamefont {Boyle}, \citenamefont
  {Filippenko}, \citenamefont {Matheson}, \citenamefont {Fruchter},
  \citenamefont {Panagia}, \citenamefont {Newberg}, \citenamefont {Couch},\
  and\ \citenamefont {Project}}]{perlmutter1999measurements}%
  \BibitemOpen
  \bibfield  {author} {\bibinfo {author} {\bibfnamefont {Perlmutter},~\bibnamefont
  {S.}}, \bibinfo {author} {\bibfnamefont {Aldering},~\bibnamefont
  {G.}}, \bibinfo {author} {\bibfnamefont {Goldhaber},~\bibnamefont {G.}},
  \bibinfo {author} {\bibfnamefont {Knop},\ \bibnamefont {R.A.}}, \bibinfo
  {author} {\bibfnamefont {Nugent},~\bibnamefont {P.}}, \bibinfo {author}
  {\bibfnamefont {Castro},\ \bibnamefont {P.G.}}, \bibinfo {author}
  {\bibfnamefont {Deustua}~\bibnamefont {S.}}, \bibinfo {author} {\bibfnamefont
  {Fabbro},~\bibnamefont {S.}}, \bibinfo {author} {\bibfnamefont
  {Goobar},~\bibnamefont {A.}}, \bibinfo {author} {\bibfnamefont {Groom},\
  \bibnamefont {D.E.}}, \bibinfo {author} {\bibfnamefont {Hook},\ \bibnamefont
  {I.M.}}, \bibinfo {author} {\bibfnamefont {Kim},\ \bibnamefont {A.G.}},
  \bibinfo {author} {\bibfnamefont {Kim},\ \bibnamefont {M.Y.}}, \bibinfo
  {author} {\bibfnamefont {Lee},\ \bibnamefont {J.C.}}, \bibinfo {author}
  {\bibfnamefont {Nunes},\ \bibnamefont {N.J.}}, \bibinfo {author}
  {\bibfnamefont {Pain},~\bibnamefont {R.}}, \bibinfo {author} {\bibfnamefont
  {Pennypacker},\ \bibnamefont {C.R.}}, \bibinfo {author} {\bibfnamefont
  {Quimby},~\bibnamefont {R.}}, \bibinfo {author} {\bibfnamefont
  {Lidman},~\bibnamefont {C.}}, \bibinfo {author} {\bibfnamefont {Ellis},\
  \bibnamefont {R.S.}}, \bibinfo {author} {\bibfnamefont {Irwin},~\bibnamefont
  {M.}}, \bibinfo {author} {\bibfnamefont {McMahon},\ \bibnamefont {R.G.}},
  \bibinfo {author} {\bibfnamefont {Ruiz‐Lapuente},~\bibnamefont {P.}},
  \bibinfo {author} {\bibfnamefont {Walton},~\bibnamefont {N.}}, \bibinfo
  {author} {\bibfnamefont {Schaefer},~\bibnamefont {B.}}, \bibinfo {author}
  {\bibfnamefont {Boyle},\ \bibnamefont {B.J.}}, \bibinfo {author}
  {\bibfnamefont {Filippenko},\ \bibnamefont {A.V.}}, \bibinfo {author}
  {\bibfnamefont {Matheson},~\bibnamefont {T.}}, \bibinfo {author}
  {\bibfnamefont {Fruchter},\ \bibnamefont {A.S.}}, \bibinfo {author}
  {\bibfnamefont {Panagia},~\bibnamefont {N.}}, \bibinfo {author} {\bibfnamefont
  {Newberg},\ \bibnamefont {H.J.M.}}, \bibinfo {author} {\bibfnamefont
  {Couch},\ \bibnamefont {W.J.}}, \ \&\ \bibinfo {author} {\bibfnamefont
  {The Supernova Cosmology}\ \bibnamefont {Project}},\ } {\emph {\bibinfo
  {title} {Measurements of $\Omega$ and $\Lambda$ from 42 High-Redshift Supernovae}}}, \href {\doibase 10.1086/307221}
  {\bibfield  {journal} {\bibinfo  {journal} {The Astrophysical Journal}\
  }\textbf {\bibinfo {volume} {517}},\ \bibinfo {pages} {565} (\bibinfo {year}
  {1999})}\BibitemShut {}%
\bibitem [{\citenamefont {Goobar}\ and\ \citenamefont
  {Leibundgut}(2011)}]{goobar2011supernova}%
  \BibitemOpen
  \bibfield  {author} {\bibinfo {author} {\bibfnamefont {Goobar},~\bibnamefont
  {A.}},\ \&\ \bibinfo {author} {\bibfnamefont {Leibundgut},~\bibnamefont
  {B.}},\ } {\emph {\bibinfo
  {title} {Supernova Cosmology: Legacy and Future}}}, \href {\doibase 10.1146/annurev-nucl-102010-130434}
  {\bibfield  {journal} {\bibinfo  {journal} {Annual Review of Nuclear and
  Particle Science}\ }\textbf {\bibinfo {volume} {61}},\ \bibinfo {pages} {251}
  (\bibinfo {year} {2011})}\BibitemShut {}%
\bibitem [{\citenamefont {Krauss}\ and\ \citenamefont
  {Chaboyer}(2003)}]{krauss2003age}%
  \BibitemOpen
  \bibfield  {author} {\bibinfo {author} {\bibfnamefont {Krauss},\ \bibnamefont
  {L.M.}},\ \&\ \bibinfo {author} {\bibfnamefont {Chaboyer},~\bibnamefont
  {B.}},\ }  {\emph {\bibinfo
  {title} {Age Estimates of Globular Clusters in the Milky Way: Constraints on Cosmology}}}, \href {\doibase 10.1126/science.1075631} {\bibfield  {journal}
  {\bibinfo  {journal} {Science}\ }\textbf {\bibinfo {volume} {299}},\ \bibinfo
  {pages} {65} (\bibinfo {year} {2003})}\BibitemShut {}%
\bibitem [{\citenamefont {Maeder}(2017{\natexlab{b}})}]{maeder2017dynamical}%
  \BibitemOpen
  \bibfield  {author} {\bibinfo {author} {\bibfnamefont {Maeder},~\bibnamefont
  {A.}},\ } {\emph {\bibinfo
  {title} {Dynamical Effects of the Scale Invariance of the Empty Space: The Fall of Dark Matter?}}} \href {\doibase 10.3847/1538-4357/aa92cc} {\bibfield  {journal}
  {\bibinfo  {journal} {The Astrophysical Journal}\ }\textbf {\bibinfo {volume}
  {849}},\ \bibinfo {pages} {158} (\bibinfo {year}
  {2017}{\natexlab{b}})}\BibitemShut {}%
\bibitem [{\citenamefont {{Guth}}(1981)}]{Guth81Inflation}%
  \BibitemOpen
  \bibfield  {author} {\bibinfo {author} {\bibfnamefont {Guth},\ \bibnamefont
  {{A.H.}}},\ } {\emph {\bibinfo
  {title} {Inflationary universe: A possible solution to the horizon and flatness problems}}}, \href {\doibase 10.1103/PhysRevD.23.347} {\bibfield  {journal}
  {\bibinfo  {journal} {\prd}\ }\textbf {\bibinfo {volume} {23}},\ \bibinfo
  {pages} {347} (\bibinfo {year} {1981})}\BibitemShut {}%
  
\bibitem [{\citenamefont {{Colin}}\ \emph {et~al.}(2018)\citenamefont
  {{Colin}}, \citenamefont {{Mohayaee}}, \citenamefont {{Rameez}},\ and\
  \citenamefont {{Sarkar}}}]{subir2018}%
  \BibitemOpen
  \bibfield  {author} {\bibinfo {author} {\bibfnamefont {Colin},~\bibnamefont
  {{J.}}}, \bibinfo {author} {\bibfnamefont {Mohayaee},~\bibnamefont {{R.}}},
  \bibinfo {author} {\bibfnamefont {Rameez},~\bibnamefont {{M.}}}, \ \&\
  \bibinfo {author} {\bibfnamefont {Sarkar},~\bibnamefont {{S.}}},\ }  {\emph {\bibinfo
  {title} {Apparent cosmic acceleration due to local bulk flow}}}, \href@noop
  {} {\bibfield  {journal} }\Eprint {http://arxiv.org/abs/1808.04597}
  {arXiv:1808.04597 (2018)} \BibitemShut {}%  
\bibitem [{\citenamefont {Copeland}\ \emph {et~al.}(2006)\citenamefont
  {Copeland}, \citenamefont {Sami},\ and\ \citenamefont
  {Tsujikawa}}]{copeland2006}%
  \BibitemOpen
  \bibfield  {author} {\bibinfo {author} {\bibfnamefont {Copeland},\ \bibnamefont
  {E.J.}}, \bibinfo {author} {\bibfnamefont {Sami},~\bibnamefont {M.}}, \
  \&\ \bibinfo {author} {\bibfnamefont {Tsujikawa},~\bibnamefont {S.}},\ }  {\emph {\bibinfo
  {title} {Dynamics of dark energy}}}, \href
  {\doibase 10.1142/S021827180600942X} {\bibfield  {journal} {\bibinfo
  {journal} {International Journal of Modern Physics D}\ }\textbf {\bibinfo
  {volume} {15}},\ \bibinfo {pages} {1753} (\bibinfo {year} {2006})}\  \BibitemShut {}%
\bibitem [{\citenamefont {Goldhaber}(2010)}]{goldhaberphoton}%
  \BibitemOpen
  \bibfield  {author} {\bibinfo {author} {\bibfnamefont {Goldhaber},~\bibnamefont
  {A.S.}},\ \&\ \bibinfo {author} {\bibfnamefont {Nieto},~\bibnamefont {M.M.}},\ }
  {\emph {\bibinfo
  {title} {Photon and graviton mass limits}}},
  \href{https://journals.aps.org/rmp/abstract/10.1103/RevModPhys.82.939} {\bibfield  {journal} {\bibinfo  {journal}
  {Review of Modern Physics}\ }\textbf {\bibinfo {volume} {82}},\ \bibinfo
  {pages} {939} (\bibinfo {year} {2010})}\BibitemShut {}%
\bibitem [{\citenamefont {Enqvist}(2008)}]{enqvist2008lemaitretolmanbondi}%
  \BibitemOpen
  \bibfield  {author} {\bibinfo {author} {\bibfnamefont {Enqvist},~\bibnamefont
  {K.}},\ } {\emph {\bibinfo
  {title} {Lemaitre–Tolman–Bondi model and accelerating expansion}}}, \href {\doibase 10.1007/s10714-007-0553-9} {\bibfield
  {journal} {\bibinfo  {journal} {General Relativity and Gravitation}\ }\textbf
  {\bibinfo {volume} {40}},\ \bibinfo {pages} {451} (\bibinfo {year}
  {2008})}\BibitemShut {}%
\bibitem [{\citenamefont {Buchert}(2000)}]{buchert2000on}%
  \BibitemOpen
  \bibfield  {author} {\bibinfo {author} {\bibfnamefont {Buchert},~\bibnamefont
  {T.}},\ } {\emph {\bibinfo
  {title} {On Average Properties of Inhomogeneous Fluids in General Relativity: Dust Cosmologies}}}, \href {\doibase 10.1023/a:1001800617177} {\bibfield  {journal}
  {\bibinfo  {journal} {General Relativity and Gravitation}\ }\textbf {\bibinfo
  {volume} {32}},\ \bibinfo {pages} {105} (\bibinfo {year} {2000})}\BibitemShut
  {}%
\bibitem [{\citenamefont {Buchert}(2001)}]{buchert2001on}%
  \BibitemOpen
  \bibfield  {author} {\bibinfo {author} {\bibfnamefont {Buchert},~\bibnamefont
  {T.}},\ }  {\emph {\bibinfo
  {title} {On Average Properties of Inhomogeneous Fluids in General Relativity: Perfect Fluid Cosmologies}}}, \href {\doibase 10.1023/a:1012061725841} {\bibfield  {journal}
  {\bibinfo  {journal} {General Relativity and Gravitation}\ }\textbf {\bibinfo
  {volume} {33}},\ \bibinfo {pages} {1381} (\bibinfo {year}
  {2001})}\BibitemShut {}%
\bibitem [{\citenamefont {Alnes}\ \emph {et~al.}(2006)\citenamefont {Alnes},
  \citenamefont {Amarzguioui},\ and\ \citenamefont
  {Gr\o{}n}}]{alnes2006inhomogeneous}%
  \BibitemOpen
  \bibfield  {author} {\bibinfo {author} {\bibfnamefont {Alnes},~\bibnamefont
  {H.}}, \bibinfo {author} {\bibfnamefont {Amarzguioui},~\bibnamefont {M.}},
  \ \&\ \bibinfo {author} {\bibfnamefont {Gr\o{}n},~\bibnamefont {O.}},\ }  {\emph {\bibinfo
  {title} {Inhomogeneous alternative to dark energy?}}} \href
  {\doibase 10.1103/PhysRevD.73.083519} {\bibfield  {journal} {\bibinfo
  {journal} {Phys. Rev. D}\ }\textbf {\bibinfo {volume} {73}},\ \bibinfo
  {pages} {083519} (\bibinfo {year} {2006})}\BibitemShut {}%
\bibitem [{\citenamefont {Hogg}\ \emph {et~al.}(2005)\citenamefont {Hogg},
  \citenamefont {Eisenstein}, \citenamefont {Blanton}, \citenamefont {Bahcall},
  \citenamefont {Brinkmann}, \citenamefont {Gunn},\ and\ \citenamefont
  {Schneider}}]{Hogg}%
  \BibitemOpen
  \bibfield  {author} {\bibinfo {author} {\bibfnamefont {Hogg},\ \bibnamefont
  {D.W.}}, \bibinfo {author} {\bibfnamefont {Eisenstein},\ \bibnamefont
  {D.J.}}, \bibinfo {author} {\bibfnamefont {Blanton},\ \bibnamefont
  {M.R.}}, \bibinfo {author} {\bibfnamefont {Bahcall},\ \bibnamefont
  {N.A.}}, \bibinfo {author} {\bibfnamefont {Brinkmann},~\bibnamefont {J.}},
  \bibinfo {author} {\bibfnamefont {Gunn},\ \bibnamefont {J.E.}}, \ \&\
  \bibinfo {author} {\bibfnamefont {Schneider},\ \bibnamefont {D.P.}},\ }  {\emph {\bibinfo
  {title} {Cosmic Homogeneity Demonstrated with Luminous Red Galaxies}}}, \href
  {http://stacks.iop.org/0004-637X/624/i=1/a=54} {\bibfield  {journal}
  {\bibinfo  {journal} {The Astrophysical Journal}\ }\textbf {\bibinfo {volume}
  {624}},\ \bibinfo {pages} {54} (\bibinfo {year} {2005})}\BibitemShut
  {}%
\bibitem [{\citenamefont {{Eisenstein}}\ \emph {et~al.}(2005)\citenamefont
  {{Eisenstein}}, \citenamefont {{Zehavi}}, \citenamefont {{Hogg}},
  \citenamefont {{Scoccimarro}}, \citenamefont {{Blanton}}, \citenamefont
  {{Nichol}}, \citenamefont {{Scranton}}, \citenamefont {{Seo}}, \citenamefont
  {{Tegmark}}, \citenamefont {{Zheng}}, \citenamefont {{Anderson}},
  \citenamefont {{Annis}}, \citenamefont {{Bahcall}}, \citenamefont
  {{Brinkmann}}, \citenamefont {{Burles}}, \citenamefont {{Castander}},
  \citenamefont {{Connolly}}, \citenamefont {{Csabai}}, \citenamefont {{Doi}},
  \citenamefont {{Fukugita}}, \citenamefont {{Frieman}}, \citenamefont
  {{Glazebrook}}, \citenamefont {{Gunn}}, \citenamefont {{Hendry}},
  \citenamefont {{Hennessy}}, \citenamefont {{Ivezi{\'c}}}, \citenamefont
  {{Kent}}, \citenamefont {{Knapp}}, \citenamefont {{Lin}}, \citenamefont
  {{Loh}}, \citenamefont {{Lupton}}, \citenamefont {{Margon}}, \citenamefont
  {{McKay}}, \citenamefont {{Meiksin}}, \citenamefont {{Munn}}, \citenamefont
  {{Pope}}, \citenamefont {{Richmond}}, \citenamefont {{Schlegel}},
  \citenamefont {{Schneider}}, \citenamefont {{Shimasaku}}, \citenamefont
  {{Stoughton}}, \citenamefont {{Strauss}}, \citenamefont {{SubbaRao}},
  \citenamefont {{Szalay}}, \citenamefont {{Szapudi}}, \citenamefont
  {{Tucker}}, \citenamefont {{Yanny}},\ and\ \citenamefont
  {{York}}}]{2005ApJ_EP}%
  \BibitemOpen
  \bibfield  {author} {\bibinfo {author} {\bibfnamefont {Eisenstein},\ \bibnamefont
  {{D.J.}}}, \bibinfo {author} {\bibfnamefont {Zehavi},~\bibnamefont
  {{I.}}}, \bibinfo {author} {\bibfnamefont {Hogg},\ \bibnamefont
  {{D.W.}}}, \bibinfo {author} {\bibfnamefont {Scoccimarro},~\bibnamefont
  {{R.}}}, \bibinfo {author} {\bibfnamefont {Blanton},\ \bibnamefont
  {{M.R.}}}, \bibinfo {author} {\bibfnamefont {Nichol},\ \bibnamefont
  {{R.C.}}}, \bibinfo {author} {\bibfnamefont {Scranton},~\bibnamefont
  {{R.}}}, \bibinfo {author} {\bibfnamefont {Seo},\ \bibnamefont
  {{H.-J.}}}, \bibinfo {author} {\bibfnamefont {Tegmark},~\bibnamefont {{M.}}},
  \bibinfo {author} {\bibfnamefont {Zheng},~\bibnamefont {{Z.}}}, \bibinfo
  {author} {\bibfnamefont {Anderson},\ \bibnamefont {{S.F.}}}, \bibinfo
  {author} {\bibfnamefont {Annis},~\bibnamefont {{J.}}}, \bibinfo {author}
  {\bibfnamefont {Bahcall},~\bibnamefont {{N.}}}, \bibinfo {author}
  {\bibfnamefont {Brinkmann},~\bibnamefont {{J.}}}, \bibinfo {author}
  {\bibfnamefont {Burles},~\bibnamefont {{S.}}}, \bibinfo {author}
  {\bibfnamefont {Castander},\ \bibnamefont {{F.J.}}}, \bibinfo {author}
  {\bibfnamefont {Connolly},~\bibnamefont {{A.}}}, \bibinfo {author}
  {\bibfnamefont {Csabai},~\bibnamefont {{I.}}}, \bibinfo {author}
  {\bibfnamefont {Doi},~\bibnamefont {{M.}}}, \bibinfo {author} {\bibfnamefont
  {Fukugita},~\bibnamefont {{M.}}}, \bibinfo {author} {\bibfnamefont {Frieman},\
  \bibnamefont {{J.A.}}}, \bibinfo {author} {\bibfnamefont {Glazebrook},~\bibnamefont
  {{K.}}}, \bibinfo {author} {\bibfnamefont {Gunn},\ \bibnamefont
  {{J.E.}}}, \bibinfo {author} {\bibfnamefont {Hendry},\ \bibnamefont
  {{J.S.}}}, \bibinfo {author} {\bibfnamefont {Hennessy},~\bibnamefont
  {{G.}}}, \bibinfo {author} {\bibfnamefont {Ivezi{\'c}},~\bibnamefont
  {{Z.}}}, \bibinfo {author} {\bibfnamefont {Kent},~\bibnamefont
  {{S.}}}, \bibinfo {author} {\bibfnamefont {Knapp},\ \bibnamefont {{G.R.}}},
  \bibinfo {author} {\bibfnamefont {Lin},~\bibnamefont {{H.}}}, \bibinfo
  {author} {\bibfnamefont {Loh},\ \bibnamefont {{Y.-S.}}}, \bibinfo {author}
  {\bibfnamefont {Lupton},\ \bibnamefont {{R.H.}}}, \bibinfo {author}
  {\bibfnamefont {Margon},~\bibnamefont {{B.}}}, \bibinfo {author}
  {\bibfnamefont {McKay},\ \bibnamefont {{T.A.}}}, \bibinfo {author}
  {\bibfnamefont {Meiksin},~\bibnamefont {{A.}}}, \bibinfo {author}
  {\bibfnamefont {Munn},\ \bibnamefont {{J.A.}}}, \bibinfo {author}
  {\bibfnamefont {Pope},~\bibnamefont {{A.}}}, \bibinfo {author} {\bibfnamefont
  {Richmond},\ \bibnamefont {{M.W.}}}, \bibinfo {author} {\bibfnamefont
  {Schlegel},~\bibnamefont {{D.}}}, \bibinfo {author} {\bibfnamefont {Schneider},\
  \bibnamefont {{D.P.}}}, \bibinfo {author} {\bibfnamefont
  {Shimasaku},~\bibnamefont {{K.}}}, \bibinfo {author} {\bibfnamefont
  {Stoughton},~\bibnamefont {{C.}}}, \bibinfo {author} {\bibfnamefont {Strauss},\
  \bibnamefont {{M.A.}}}, \bibinfo {author} {\bibfnamefont {SubbaRao},~\bibnamefont
  {{M.}}}, \bibinfo {author} {\bibfnamefont {Szalay},\ \bibnamefont
  {{A.S.}}}, \bibinfo {author} {\bibfnamefont {Szapudi},~\bibnamefont {{I.}}},
  \bibinfo {author} {\bibfnamefont {Tucker},\ \bibnamefont {{D.L.}}}, \bibinfo
  {author} {\bibfnamefont {Yanny},~\bibnamefont {{B.}}}, \ \&\ \bibinfo
  {author} {\bibfnamefont {York},\ \bibnamefont {{D.G.}}},\ }  {\emph {\bibinfo
  {title} {Detection of the Baryon Acoustic Peak in the Large-Scale Correlation Function of SDSS Luminous Red Galaxies}}}, \href {\doibase
  10.1086/466512} {\bibfield  {journal} {\bibinfo  {journal} {\apj}\ }\textbf
  {\bibinfo {volume} {633}},\ \bibinfo {pages} {560} (\bibinfo {year}
  {2005})}\BibitemShut {}%
\bibitem [{\citenamefont {He\ss}\ and\ \citenamefont {Kitaura}(2016)}]{Hess}%
  \BibitemOpen
  \bibfield  {author} {\bibinfo {author} {\bibfnamefont {He\ss},~\bibnamefont
  {S.}},\ \&\ \bibinfo {author} {\bibfnamefont {Kitaura},\ \bibnamefont
  {F.-S.}},\ }  {\emph {\bibinfo
  {title} {Cosmic flows and the expansion of the local Universe from non-linear phase–space reconstructions}}}, \href {\doibase 10.1093/mnras/stv2928} {\bibfield  {journal}
  {\bibinfo  {journal} {Monthly Notices of the Royal Astronomical Society}\
  }\textbf {\bibinfo {volume} {456}},\ \bibinfo {pages} {4247} (\bibinfo {year}
  {2016})}\BibitemShut {}%
\bibitem [{\citenamefont {Peccei}\ and\ \citenamefont
  {Quinn}(1977)}]{peccei1977cp}%
  \BibitemOpen
  \bibfield  {author} {\bibinfo {author} {\bibfnamefont {Peccei},~\bibnamefont
  {R.}},\ \&\ \bibinfo {author} {\bibfnamefont {Quinn},~\bibnamefont {H.}},\
  } {\emph {\bibinfo
  {title} {CP conservation in the presence of pseudoparticles}}}, \href {https://journals.aps.org/prl/abstract/10.1103/PhysRevLett.38.1440} {\bibfield  {journal} {\bibinfo  {journal}
  {Physical Review Letters}\ }\textbf {\bibinfo {volume} {38}},\ \bibinfo
  {pages} {1440} (\bibinfo {year} {1977})}\BibitemShut {}%
\bibitem [{\citenamefont {Wilczek}(1978)}]{wilczek1978problem}%
  \BibitemOpen
  \bibfield  {author} {\bibinfo {author} {\bibfnamefont {Wilczek},~\bibnamefont
  {F.}},\ } {\emph {\bibinfo
  {title} {Problem of Strong $P$ and $T$ Invariance in the Presence of Instantons}}}, \href {\doibase 10.1103/physrevlett.40.279} {\bibfield
  {journal} {\bibinfo  {journal} {Phys Rev Lett}\ }\textbf {\bibinfo {volume}
  {40}},\ \bibinfo {pages} {279} (\bibinfo {year} {1978})}\BibitemShut
  {}%
\bibitem [{\citenamefont {Jaeckel}\ and\ \citenamefont
  {Ringwald}(2010)}]{jaeckel2010the}%
  \BibitemOpen
  \bibfield  {author} {\bibinfo {author} {\bibfnamefont {Jaeckel},~\bibnamefont
  {J.}},\ \&\ \bibinfo {author} {\bibfnamefont {Ringwald},~\bibnamefont
  {A.}},\ } {\emph {\bibinfo
  {title} {The Low-Energy Frontier of Particle Physics}}}, \href {\doibase 10.1146/annurev.nucl.012809.104433} {\bibfield
   {journal} {\bibinfo  {journal} {Annu. Rev. Nucl. Part. Sci.}\ }\textbf
  {\bibinfo {volume} {60}},\ \bibinfo {pages} {405} (\bibinfo {year}
  {2010})}\BibitemShut {}%
\bibitem [{\citenamefont {Dine}\ and\ \citenamefont
  {Fischler}(1983)}]{DINE1983137}%
  \BibitemOpen
  \bibfield  {author} {\bibinfo {author} {\bibfnamefont {Dine},~\bibnamefont
  {M.}},\ \&\ \bibinfo {author} {\bibfnamefont {Fischler},~\bibnamefont
  {W.}},\ } {\emph {\bibinfo
  {title} {The not-so-harmless axion}}}, \href {https://doi.org/10.1016/0370-2693(83)90639-1}
  {\bibfield  {journal} {\bibinfo  {journal} {Physics Letters B}\ }\textbf
  {\bibinfo {volume} {120}},\ \bibinfo {pages} {137 } (\bibinfo {year}
  {1983})}\BibitemShut {}%
\bibitem [{\citenamefont {Abbott}\ and\ \citenamefont
  {Sikivie}(1983)}]{ABBOTT1983133}%
  \BibitemOpen
  \bibfield  {author} {\bibinfo {author} {\bibfnamefont {Abbott},~\bibnamefont
  {L.}},\ \&\ \bibinfo {author} {\bibfnamefont {Sikivie},~\bibnamefont
  {P.}},\ } {\emph {\bibinfo
  {title} {A cosmological bound on the invisible axion}}}, \href {https://doi.org/10.1016/0370-2693(83)90638-X}
  {\bibfield  {journal} {\bibinfo  {journal} {Physics Letters B}\ }\textbf
  {\bibinfo {volume} {120}},\ \bibinfo {pages} {133 } (\bibinfo {year}
  {1983})}\BibitemShut {}%
\bibitem [{\citenamefont {Preskill}\ \emph {et~al.}(1983)\citenamefont
  {Preskill}, \citenamefont {Wise},\ and\ \citenamefont
  {Wilczek}}]{PRESKILL1983127}%
  \BibitemOpen
  \bibfield  {author} {\bibinfo {author} {\bibfnamefont {Preskill},~\bibnamefont
  {J.}}, \bibinfo {author} {\bibfnamefont {Wise},\ \bibnamefont {M.B.}},
  \ \&\ \bibinfo {author} {\bibfnamefont {Wilczek},~\bibnamefont {F.}},\ }  {\emph {\bibinfo
  {title} {Cosmology of the invisible axion}}}, \href
  {https://doi.org/10.1016/0370-2693(83)90637-8} {\bibfield  {journal}
  {\bibinfo  {journal} {Physics Letters B}\ }\textbf {\bibinfo {volume}
  {120}},\ \bibinfo {pages} {127 } (\bibinfo {year} {1983})}\BibitemShut
  {}%
\bibitem [{\citenamefont {Schive}\ \emph {et~al.}(2014)\citenamefont {Schive},
  \citenamefont {Chiueh},\ and\ \citenamefont {Broadhurst}}]{schive2014cosmic}%
  \BibitemOpen
  \bibfield  {author} {\bibinfo {author} {\bibfnamefont {Schive},\ \bibnamefont
  {H.-Y.}}, \bibinfo {author} {\bibfnamefont {Chiueh},~\bibnamefont {T.}}, \
  \&\ \bibinfo {author} {\bibfnamefont {Broadhurst},~\bibnamefont {T.}},\
  }  {\emph {\bibinfo
  {title} {Cosmic structure as the quantum interference of a coherent dark wave}}}, \href {\doibase 10.1038/NPHYS2996} {\bibfield  {journal} {\bibinfo
  {journal} {Nature Physics}\ }\textbf {\bibinfo {volume} {10}},\ \bibinfo
  {pages} {496} (\bibinfo {year} {2014})}\BibitemShut {}%
\bibitem [{\citenamefont {Pitaevskii}\ and\ \citenamefont
  {Stringari}(2016)}]{bec}%
  \BibitemOpen
  \bibfield  {author} {\bibinfo {author} {\bibfnamefont {Pitaevskii},~\bibnamefont
  {L.}},\ \&\ \bibinfo {author} {\bibfnamefont {Stringari},~\bibnamefont
  {S.}},\ }\href@noop {} {\emph {\bibinfo {title} {Bose-Einstein
  Condensation and Superfluidity}}}\ (\bibinfo  {publisher} {Oxford University
  Press},\ \bibinfo {year} {2016})\BibitemShut {}%
\bibitem [{\citenamefont {Patrignani}\ \citenamefont
  {\it et al.}(2016)}]{1674-1137-40-10-100001}%
  \BibitemOpen
  \bibfield  {author} {\bibinfo {author} {\bibfnamefont {Patrignani},~\bibnamefont
  {C.}},\ \&\ \bibinfo {author} { \bibnamefont
  {Particle Data Group}},\ } {\emph {\bibinfo
  {title} {Review of Particle Physics}}}, \href {http://stacks.iop.org/1674-1137/40/i=10/a=100001}
  {\bibfield  {journal} {\bibinfo  {journal} {Chinese Physics C}\ }\textbf
  {\bibinfo {volume} {40}},\ \bibinfo {pages} {100001} (\bibinfo {year}
  {2016})}\BibitemShut {}%
\bibitem [{\citenamefont {{Wright}}(2004)}]{2004mmu..symp..291W}%
  \BibitemOpen
  \bibfield  {author} {\bibinfo {author} {\bibfnamefont {Wright},\ \bibnamefont
  {{E.L.}}},\ } {\emph {\bibinfo
  {title} {Theoretical Overview of Cosmic Microwave Background Anisotropy}}}, \href@noop {} \Eprint {http://arxiv.org/abs/astro-ph/0305591 (2004)}
  {astro-ph/0305591} \BibitemShut {}%
\bibitem [{\citenamefont {De~Martino}\ \emph {et~al.}(2017)\citenamefont
  {De~Martino}, \citenamefont {Broadhurst}, \citenamefont {Tye}, \citenamefont
  {Chiueh}, \citenamefont {Schive},\ and\ \citenamefont
  {Lazkoz}}]{PhysRevLett.119.221103}%
  \BibitemOpen
  \bibfield  {author} {\bibinfo {author} {\bibfnamefont {De~Martino},~\bibnamefont
  {I.}}, \bibinfo {author} {\bibfnamefont {Broadhurst},~\bibnamefont
  {T.}}, \bibinfo {author} {\bibfnamefont {Tye},\ \bibnamefont
  {S.-H.H.}}, \bibinfo {author} {\bibfnamefont {Chiueh},~\bibnamefont {T.}},
  \bibinfo {author} {\bibfnamefont {Schive},\ \bibnamefont {H.-Y.}}, \ \&\
  \bibinfo {author} {\bibfnamefont {Lazkoz}~\bibnamefont {R.}},\ } {\emph {\bibinfo
  {title} {Recognizing Axionic Dark Matter by Compton and de Broglie Scale Modulation of Pulsar Timing}}}, \href
  {\doibase 10.1103/PhysRevLett.119.221103} {\bibfield  {journal} {\bibinfo
  {journal} {Phys. Rev. Lett.}\ }\textbf {\bibinfo {volume} {119}},\ \bibinfo
  {pages} {221103} (\bibinfo {year} {2017})}\BibitemShut {}%
\bibitem [{\citenamefont {{Choi}}\ \emph {et~al.}(2018)\citenamefont {{Choi}},
  \citenamefont {{Kim}},\ and\ \citenamefont
  {{Sekiguchi}}}]{2018arXiv180207269C}%
  \BibitemOpen
  \bibfield  {author} {\bibinfo {author} {\bibfnamefont {Choi},~\bibnamefont
  {{K.}}}, \bibinfo {author} {\bibfnamefont {Kim},~\bibnamefont {{H.}}}, \
  \&\ \bibinfo {author} {\bibfnamefont {Sekiguchi},~\bibnamefont {{T.}}},\
  }  {\emph {\bibinfo
  {title} {Late-Time Magnetogenesis Driven by Axionlike Particle Dark Matter and a Dark Photon}}}, \href
  {\doibase 10.1103/PhysRevLett.121.031102} {\bibfield  {journal} {\bibinfo
  {journal} {Phys. Rev. Lett.}\ }\textbf {\bibinfo {volume} {121}},\ \bibinfo
  {pages} {031102} (\bibinfo {year} {2018})}\BibitemShut {}%

\bibitem [{\citenamefont {Miniati}\ \emph {et~al.}(2018)\citenamefont
  {Miniati}, \citenamefont {Gregori}, \citenamefont {Reville},\ and\
  \citenamefont {Sarkar}}]{PhysRevLett.121.021301}%
  \BibitemOpen
  \bibfield  {author} {\bibinfo {author} {\bibfnamefont {Miniati},~\bibnamefont
  {F.}}, \bibinfo {author} {\bibfnamefont {Gregori},~\bibnamefont {G.}},
  \bibinfo {author} {\bibfnamefont {Reville},~\bibnamefont {B.}}, \ \&\
  \bibinfo {author} {\bibfnamefont {Sarkar},~\bibnamefont {S.}},\ } {\emph {\bibinfo
  {title} {Axion-Driven Cosmic Magnetogenesis during the QCD Crossover}}}, \href
  {\doibase 10.1103/PhysRevLett.121.021301} {\bibfield  {journal} {\bibinfo
  {journal} {Phys. Rev. Lett.}\ }\textbf {\bibinfo {volume} {121}},\ \bibinfo
  {pages} {021301} (\bibinfo {year} {2018})}\BibitemShut {}%

\end{thebibliography}
\end{document}